\documentclass[aps,prx,twocolumn,english,superscriptaddress,longbibliography]{revtex4-1}
\usepackage[latin9]{inputenc}
\usepackage{amsmath}
\usepackage{amsfonts} 
\usepackage{babel}
\usepackage{bm}
\usepackage{color}
\usepackage{dcolumn}
\usepackage{graphics}
\usepackage{graphicx}
\usepackage{helvet}
\usepackage{longtable}
\usepackage{subfigure}
\usepackage{xspace}
\usepackage[colorlinks = true,
            linkcolor = blue,
            urlcolor  = blue,
            citecolor = blue,
            anchorcolor = blue]{hyperref}

\makeatletter

\DeclareGraphicsExtensions{.pdf, .png, .jpg, .eps}

\makeatother

\begin{document}
\title{Revealing three-dimensional quantum criticality by Sr-substitution 
in Han Purple}

\author{Stephan Allenspach}
\affiliation{Quantum Criticality and Dynamics Group, Paul Scherrer Institute, 
CH-5232 Villigen-PSI, Switzerland.}
\affiliation{Department of Quantum Matter Physics, University of Geneva, 
CH-1211 Geneva, Switzerland.}

\author{Pascal Puphal}
\affiliation{Laboratory for Multiscale Materials Experiments, Paul Scherrer 
Institute, CH-5232 Villigen-PSI, Switzerland.}
\affiliation{Physikalisches Institut, Goethe-University Frankfurt, 60438 
Frankfurt am Main, Germany.}
\affiliation{Max Planck Institute for Solid-State Research, Heisenbergstrasse 
1, 70569 Stuttgart, Germany.}

\author{Joosep Link}
\affiliation{National Institute of Chemical Physics and Biophysics, 12618 
Tallinn, Estonia.}

\author{Ivo Heinmaa}
\affiliation{National Institute of Chemical Physics and Biophysics, 12618 
Tallinn, Estonia.}

\author{Ekaterina Pomjakushina}
\affiliation{Laboratory for Multiscale Materials Experiments, Paul Scherrer 
Institute, CH-5232 Villigen-PSI, Switzerland.}

\author{Cornelius Krellner}
\affiliation{Physikalisches Institut, Goethe-University Frankfurt, 60438 
Frankfurt am Main, Germany.}

\author{Jakob Lass}
\affiliation{Laboratory for Neutron Scattering and Imaging, Paul Scherrer 
Institute, CH-5232 Villigen-PSI, Switzerland.}
\affiliation{Nanoscience Center, Niels Bohr Institute, University of 
Copenhagen, 2100 Copenhagen, Denmark.}

\author{Gregory S. Tucker}
\affiliation{Laboratory for Neutron Scattering and Imaging, Paul Scherrer
Institute, CH-5232 Villigen-PSI, Switzerland.}
\affiliation{Institute of Physics, Ecole Polytechnique F\'{e}d\'{e}rale de 
Lausanne (EPFL), CH-1015 Lausanne, Switzerland.}

\author{Christof Niedermayer}
\affiliation{Laboratory for Neutron Scattering and Imaging, Paul Scherrer
Institute, CH-5232 Villigen-PSI, Switzerland.}

\author{Shusaku Imajo}
\affiliation{International MegaGauss Science Laboratory, Institute for Solid 
State Physics, The University of Tokyo, Kashiwa, Chiba 277-8581, Japan.}

\author{Yoshimitsu Kohama}
\affiliation{International MegaGauss Science Laboratory, Institute for Solid 
State Physics, The University of Tokyo, Kashiwa, Chiba 277-8581, Japan.}

\author{Koichi Kindo}
\affiliation{International MegaGauss Science Laboratory, Institute for Solid 
State Physics, The University of Tokyo, Kashiwa, Chiba 277-8581, Japan.}

\author{Steffen Kr\"{a}mer}
\affiliation{Laboratoire National des Champs Magn\'{e}tiques Intenses, 
LNCMI-CNRS (UPR3228), EMFL, Universit\'e \\ Grenoble Alpes, UPS and INSA 
Toulouse, Bo\^{i}te Postale 166, 38042 Grenoble Cedex 9, France.}

\author{Mladen Horvati{\'{c}}}
\affiliation{Laboratoire National des Champs Magn\'{e}tiques Intenses, 
LNCMI-CNRS (UPR3228), EMFL, Universit\'e \\ Grenoble Alpes, UPS and INSA 
Toulouse, Bo\^{i}te Postale 166, 38042 Grenoble Cedex 9, France.}

\author{Marcelo Jaime}
\affiliation{MPA-MAGLAB, Los Alamos National Laboratory, Los Alamos, New Mexico 87545, 
USA.}

\author{Alexander Madsen}
\affiliation{Quantum Criticality and Dynamics Group, Paul Scherrer Institute, 
CH-5232 Villigen-PSI, Switzerland.}
\affiliation{Institute of Computational Science, Universit\`a della Svizzera 
italiana, CH-6900 Lugano, Switzerland.}

\author{Antonietta Mira}
\affiliation{Institute of Computational Science, Universit\`a della Svizzera 
italiana, CH-6900 Lugano, Switzerland.}
\affiliation{Dipartimento di Scienza e Alta Tecnologia, Universit\`a degli 
Studi dell'Insubria, 2210 Como, Italy.}

\author{Nicolas Laflorencie}
\affiliation{Laboratoire de Physique Th\'{e}orique, CNRS and Universit\'{e} 
de Toulouse, 31062 Toulouse, France.}

\author{Fr\'{e}d\'{e}ric Mila}
\affiliation{Institute of Physics, Ecole Polytechnique F\'{e}d\'{e}rale de 
Lausanne (EPFL), CH-1015 Lausanne, Switzerland.}

\author{Bruce Normand}
\affiliation{Quantum Criticality and Dynamics Group, Paul Scherrer Institute, 
CH-5232 Villigen-PSI, Switzerland.}
\affiliation{Institute of Physics, Ecole Polytechnique F\'{e}d\'{e}rale de 
Lausanne (EPFL), CH-1015 Lausanne, Switzerland.}
\affiliation{Lehrstuhl f\"ur Theoretische Physik I, Technische Universit\"at 
Dortmund, Otto-Hahn-Strasse 4, 44221 Dortmund, Germany}

\author{Christian R\"{u}egg}
\affiliation{Quantum Criticality and Dynamics Group, Paul Scherrer Institute, 
CH-5232 Villigen-PSI, Switzerland.}
\affiliation{Department of Quantum Matter Physics, University of Geneva, 
CH-1211 Geneva, Switzerland.}
\affiliation{Institute of Physics, Ecole Polytechnique F\'{e}d\'{e}rale de 
Lausanne (EPFL), CH-1015 Lausanne, Switzerland.}
\affiliation{Institute for Quantum Electronics, ETH Z\"urich, CH-8093 
H\"onggerberg, Switzerland.}

\author{Raivo Stern}
\affiliation{National Institute of Chemical Physics and Biophysics, 12618 
Tallinn, Estonia.}

\author{Franziska Weickert}
\affiliation{National High Magnetic Field Laboratory, Florida State University, 
Tallahassee, Florida 32310, USA.}

\begin{abstract}
Classical and quantum phase transitions (QPTs), with their accompanying 
concepts of criticality and universality, are a cornerstone of statistical 
thermodynamics. An excellent example of a controlled QPT is the field-induced 
ordering of a gapped quantum magnet. Although numerous ``quasi-one-dimensional''
coupled spin-chain and -ladder materials are known whose ordering transition 
is three-dimensional (3D), quasi-2D systems are special for multiple reasons. 
Motivated by the ancient pigment Han Purple (BaCuSi$_{2}$O$_{6}$), a quasi-2D 
material displaying anomalous critical properties, we present a complete 
analysis of Ba$_{0.9}$Sr$_{0.1}$CuSi$_{2}$O$_{6}$. We measure the zero-field 
magnetic excitations by neutron spectroscopy and deduce the spin Hamiltonian. 
We probe the field-induced transition by combining magnetization, 
specific-heat, torque, and magnetocalorimetric measurements with nuclear 
magnetic resonance studies near the QPT. By a Bayesian statistical analysis 
and large-scale Quantum Monte Carlo simulations, we demonstrate unambiguously 
that observable 3D quantum critical scaling is restored by the structural 
simplification arising from light Sr-substitution in Han Purple.
\end{abstract}

\maketitle

\section{Introduction}
\label{sintro}

At a continuous classical or quantum phase transition (QPT), characteristic 
energy scales vanish, characteristic (``correlation'') lengths diverge and 
hence the properties of the system are dictated only by global and 
scale-invariant quantities \cite{Zinn-Justin2002}. The critical properties 
at the transition then depend only on factors such as the dimensionality of 
the system, the symmetry group of the order parameter, and in some cases on 
topological criteria. These fundamental factors are not all independent, but 
are all discrete, and as a result phase transitions can be categorized by 
their ``universality class.'' 

An instructive example of the effects of dimensionality is found for systems 
where the U(1) symmetry of the order parameter is broken \cite{Pitaevskii2016}.
For free bosons, the symmetry-broken phase is the Bose-Einstein condensate 
(BEC) \cite{Bose1924,Einstein1924} and, while most familiar in three dimensions 
(3D), this transition can in fact be found in systems whose effective dimension 
is any real number $d > 2$. However, strictly in 2D the physics is quite 
different, dependent on the binding of point vortices, and the system displays 
the Berezinskii-Kosterlitz-Thouless (BKT) transition \cite{Berezinskii1971,
Kosterlitz1973}. In real materials, the issue of system dimensionality is 
complicated by the fact that the coupling of different subsystems (such as 
clusters, chains, or planes) may occur on different low energy scales that are 
only revealed as the temperature is lowered. Strict 1D or 2D behavior is 
avoided because, as the critical point is approached on the low-dimensional 
subsystem, the growing correlation length causes the subsystems to become 
entangled and the critical properties are expected to be those associated 
with a 3D universality class \cite{Giamarchi2003,Sachdev2011}.

Dimerized quantum spin systems, composed of strongly coupled pairs of spin-1/2 
ions, constitute a particularly valuable class of materials for the study of 
quantum criticality. For sufficiently strong dimerization, the zero-field 
ground state of local singlets is always gapped, with the excitations taking 
the form of propagating local triplet quasiparticles (``triplons''). In an 
applied magnetic field, this system undergoes a QPT at which the gap is 
closed and the new ground state has field-induced transverse magnetic order, 
which can be described as a triplon condensate \cite{Nikuni2000,Rice2002,
Giamarchi2008,Zapf2014}. This is a 3D QPT from a quantum disordered phase 
with U(1) (or XY) symmetry to a BEC phase, whose emerging long-range magnetic 
order breaks the U(1) symmetry. This QPT has been studied extensively in 
materials where the inter-dimer coupling is similar in all three spatial 
dimensions \cite{Rueegg2003,Giamarchi2008}. In gapped quasi-1D materials 
(dimerized chains, Haldane chains, two-leg ladders), fields exceeding the gap 
reveal the physics of the Tomonaga-Luttinger liquid at higher temperatures, 
but 3D scaling behavior is restored in the quantum critical regime around 
the QPT \cite{Klanjsek2008,Thielemann2008,Schmidiger2012,Mukhopadhyay2012,
Blinder2017,Tsirlin2019}. However, 2D systems with continuous symmetry are 
special because of the Mermin-Wagner theorem \cite{Mermin1966}, and 
extra-special because of BKT physics \cite{Berezinskii1971,Kosterlitz1973}, 
which has recently been argued to give rise to behavior at the dimensional 
crossover that is radically different from the 0D and 1D cases 
\cite{Furuya2016}. Thus the question arises as to whether the 
``conventional'' emergence of 3D physics can really be expected 
at the field-induced QPT in a quasi-2D material. 

Unfortunately the list of candidate quasi-2D spin-dimer compounds suitable 
for answering this question is rather short. One well-studied material is 
SrCu$_2$(BO$_3$)$_2$ \cite{Kageyama1999}, which displays a wealth of 
fascinating physics as a consequence of its ideally frustrated 
Shastry-Sutherland geometry \cite{Sutherland1981}. While this includes 
magnetization plateaus \cite{Momoi2000,Kodama2002,Takigawa2013} and 
topological phases \cite{Romhanyi2015} in an applied field, first-order QPTs 
to plaquette \cite{Corboz2013,Corboz2014} and antiferromagnetic (AF) phases 
under applied pressure \cite{Waki2007,Takigawa2010,Haravifard2012,Zayed2017,
Sakurai2018}, proliferating bound-state excitations, and anomalous 
thermodynamics \cite{Wietek2019}, it does not include the type of physics 
we discuss. The compound (C$_4$H$_{12}$N$_2$)Cu$_2$Cl$_6$ was also an early 
candidate due to its well-separated planes of Cl-coordinated Cu dimers, but 
was found to possess complicated and frustrated 3D coupling \cite{Stone2001} 
and to display an unexpected breakdown of triplon excitations due to strong 
mutual scattering \cite{Stone2006}. The chromate compounds Ba$_3$Cr$_2$O$_8$ 
\cite{Nakajima2006,Kofu2009,Aczel2009a} and Sr$_3$Cr$_2$O$_8$ \cite{Aczel2009b,
Nomura2020} form $S = 1/2$ dimer units in a triangular geometry and show 3D 
scaling at the field-induced QPT, but structural distortion and magnetic 
frustration effects leave their dominant dimensionality (2D or 1D) unconfirmed. 

A further prominent material, which does actually realize 2D square 
lattices of dimers and a field-induced triplon condensation at $\mu_0 
H_{c1} \simeq 23.5$ T, is the ancient pigment Han Purple (BaCuSi$_{2}$O$_{6}$) 
\cite{Jaime2004,Sebastian2005}. In this compound the stacked dimer bilayers 
have a body-centered offset, leading to the expectation of an exact frustration 
of AF coupling between successive bilayers, and further to the interpretation 
of thermodynamic measurements showing apparent 2D scaling close to the QPT 
\cite{Sebastian2006} as a frustration-induced ``dimensional reduction'' quite 
different from 3D critical properties. However, it was shown subsequently that 
the weakly orthorhombic structure of this material below 90~K \cite{Samulon2006,
Sparta2006,Stern2014} contains three types of structurally \cite{Sheptyakov2012}
and magnetically \cite{Horvatic2005,Rueegg2007,Kraemer2007} inequivalent 
bilayers, and later that the inter-bilayer coupling is not in fact frustrated 
\cite{Mazurenko2014,Allenspach2020}. These properties lead ultimately to a 
regime of anomalous effective scaling in the experimentally accessible 
parameter space \cite{Allenspach2020}, with the 3D critical behavior 
remaining unobservable. 

\begin{figure*}[t]
\begin{centering}
\includegraphics[width=\textwidth]{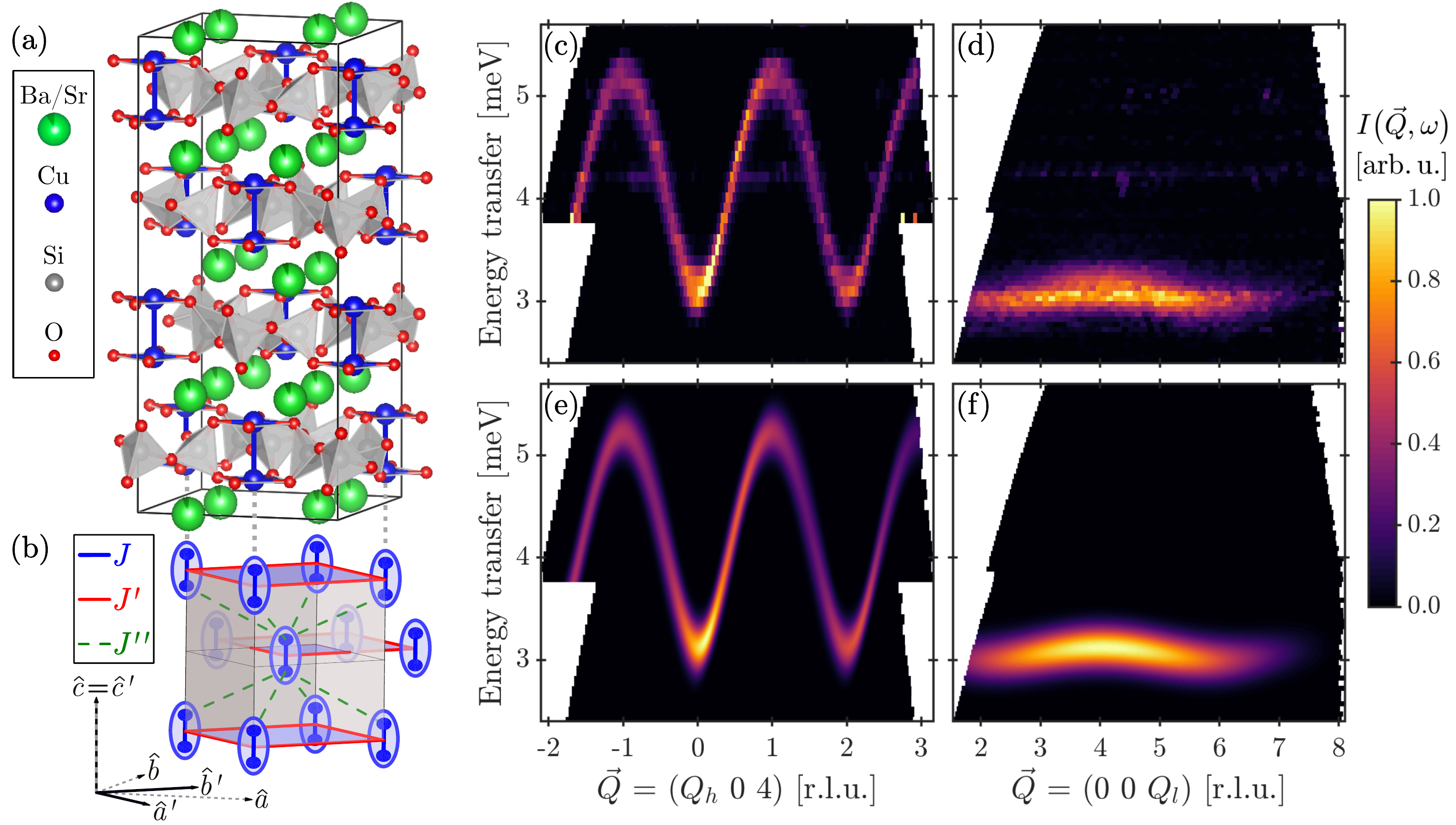} 
\end{centering}
\caption{{\bf Structure and inelastic neutron spectrum of 
Ba$_{0.9}$Sr$_{0.1}$CuSi$_{2}$O$_{6}$.} (a) Tetragonal crystallographic 
structure of Ba$_{0.9}$Sr$_{0.1}$CuSi$_{2}$O$_{6}$ \cite{Puphal2016}. (b) 
Representation of the minimal magnetic unit cell, indicating the relevant 
Heisenberg interaction parameters between dimers on the two bilayers, 
which are structurally equivalent with body-centered geometry. $({\hat a}, 
{\hat b}, {\hat c})$ and $({\hat a}', {\hat b}', {\hat c}')$ denote 
respectively the axes of the crystallographic and minimal magnetic unit 
cells \cite{Allenspach2020}. Dotted grey lines indicate the correspondence 
between Cu$^{2+}$ ion positions in the two cells. (c)-(d) Neutron scattering 
spectra measured on CAMEA at 1.5~K for two high-symmetry ${\vec Q}$-space 
directions; ${\vec Q}$ is indexed in reciprocal lattice units (r.l.u.)~of 
the crystallographic unit cell. (e)-(f) Spectra modelled using the interaction 
parameters and linewidths extracted from the data collected on both CAMEA 
and TASP (Fig.~\ref{fig2}).}
\label{fig1}
\end{figure*}

At room temperature, BaCuSi$_2$O$_6$ has a tetragonal structure (space group 
I4$_{1}$/acd) in which all dimer bilayers are equivalent. Thus a possible 
route to the ideal quasi-2D material would be to suppress the orthorhombic 
transition. In pursuing this program, a number of stoichiometric 
substitutions have been tested on the non-magnetic Ba site, and it was 
reported \cite{Puphal2016} that even a 5\% Sr substitution is sufficient to 
stabilize the tetragonal structure, represented in Fig.~\ref{fig1}(a), down 
to the lowest temperatures. However, it is a significant challenge to grow 
single crystals of the substituted material \cite{Well2016}, and particularly 
to grow large single crystals suitable for inelastic neutron scattering (INS) 
experiments \cite{Puphal2019}. We have solved these problems, enabling the 
detailed study of the ground state, excitations, and field-temperature phase 
diagram of Ba$_{0.9}$Sr$_{0.1}$CuSi$_{2}$O$_{6}$ that we present.

\begin{figure*}[t]
\noindent
\begin{centering}
\includegraphics[width=\textwidth]{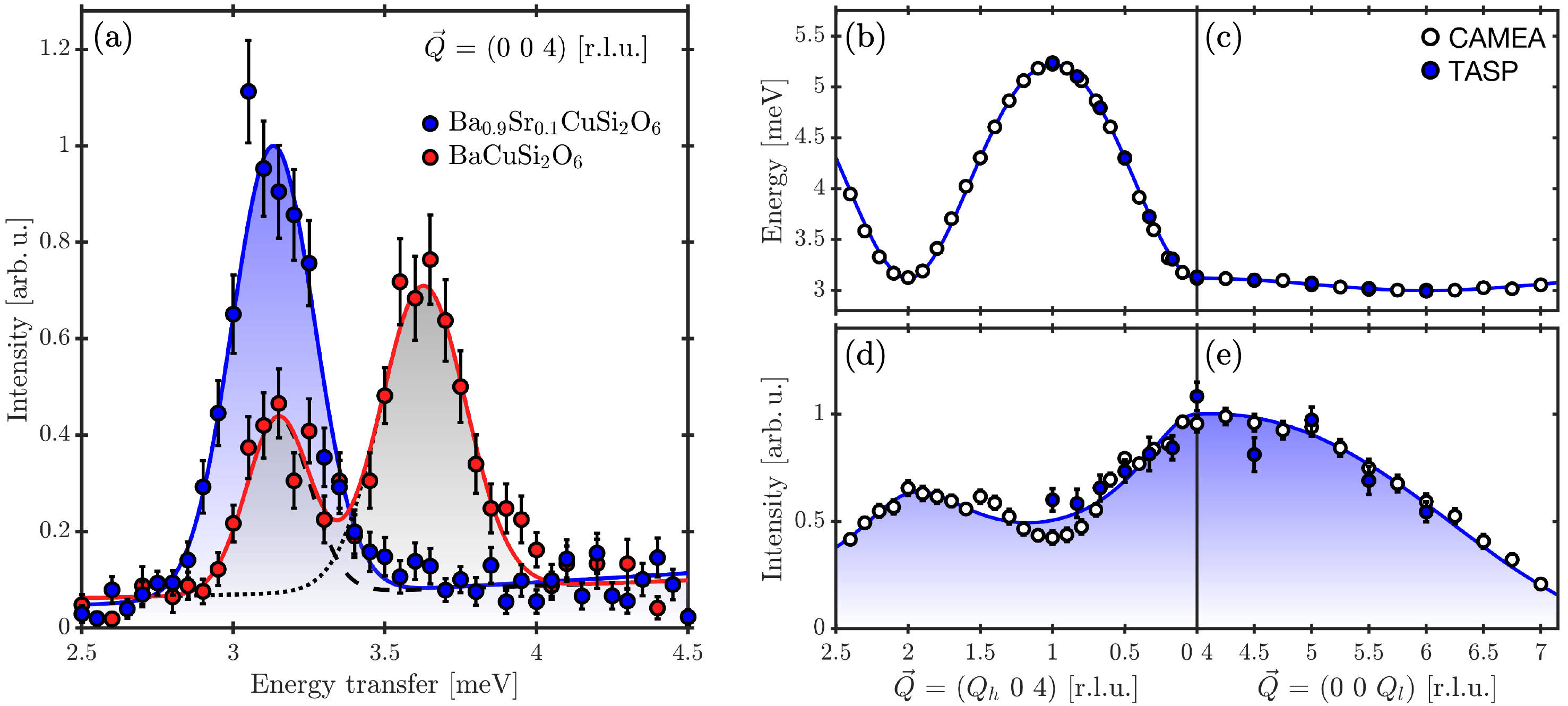} 
\end{centering}
\caption{{\bf Inelastic neutron scattering analysis.} (a) Scattered intensity 
as a function of energy transfer, measured on TASP at ${\vec Q} = (0 \; 0 
\; 4)$ r.l.u. for Ba$_{0.9}$Sr$_{0.1}$CuSi$_{2}$O$_{6}$ at 1.5~K (blue symbols) 
and BaCuSi$_{2}$O$_{6}$ at 1.6~K (red). Shaded regions represent Gaussian fits 
used to extract the triplon energy, linewidth, and intensity. (b)-(e) Multiple 
Ba$_{0.9}$Sr$_{0.1}$CuSi$_{2}$O$_{6}$ energy scans, of the type shown in panel 
(a), were analyzed for ${\vec Q}$-vectors covering two high-symmetry directions 
in reciprocal space. (b)-(c) Triplon mode energies obtained from data gathered 
on TASP (blue symbols) and CAMEA (open). (d)-(e) Corresponding scattered 
intensities. Solid lines are fits obtained from the spin Hamiltonian 
represented in Fig.~\ref{fig1}(b) with the optimal parameters given 
in the text.} 
\label{fig2}
\end{figure*}

The structure of our contribution is as follows. In Sec.~\ref{smm} we detail 
our crystal-growth procedures, experimental methods, and numerical simulations. 
In Sec.~\ref{sins} we report INS experiments at zero magnetic field, which 
demonstrate that this compound contains only one type of dimer bilayer, 
and hence one triplon excitation branch. We determine the minimal spin 
Hamiltonian, showing that it contains only the three Heisenberg interactions 
of Fig.~\ref{fig1}(b), which form a clear hierarchy of intradimer ($J$), 
intra-bilayer ($J^{\prime}$), and inter-bilayer ($J^{\prime\prime}$) coupling 
strengths. Turning to high magnetic fields, in Sec.~\ref{spb} we perform 
magnetization, magnetic torque, specific-heat, magnetocaloric effect (MCE), 
and nuclear magnetic resonance (NMR) measurements to obtain a comprehensive 
experimental picture of the field-temperature phase boundary to the 
magnetically ordered state. In Sec.~\ref{sce} we combine state-of-the-art 
statistical data analysis and Quantum Monte Carlo (QMC) simulations based 
on the INS parameters to interpret our results as demonstrating complete 
consistency with 3D scaling in the quantum critical region. In Sec.~\ref{sd} 
we discuss the context of these results, particularly regarding possible 
magnetic disorder effects, and conclude in Sec.~\ref{sc} that the structural 
consequences of weak chemical substitution by non-magnetic ions are sufficient 
to realize the simplicity and elegance of the model originally envisaged for 
Han Purple, namely a quasi-2D spin-dimer material displaying observable 3D 
quantum critical behavior.

\section{Material and Methods}
\label{smm}

\subsection{Crystal growth}

Single-crystal growth of BaCuSi$_{2}$O$_{6}$ poses a challenge because the 
starting CuO component decays before the reagents and compound melt. This 
decay has been suppressed by increasing the oxygen partial pressure in 
optical floating-zone growth \cite{Jaime2004} and by using oxygen-donating 
LiBO$_{2}$ flux \cite{Sebastian2006}. However, the latter method cannot be 
applied to the substituted system, Ba$_{0.9}$Sr$_{0.1}$CuSi$_{2}$O$_{6}$, 
because the metaborate flux reacts with the SrO reagent \cite{Well2016}.
Thus two methods were employed to grow the single crystals used in the 
present work. All of the high-field experiments were performed using small 
single crystals (sizes of order 10 mg) grown from the congruent melt in a tube 
furnace with an oxygen flow at a pressure under 1 bar \cite{Well2016}. The 
INS experiments were performed using one rod-shaped single crystal of mass
1.3 g grown by a floating-zone method. While this method was developed 
\cite{Puphal2019} for BaCuSi$_{2}$O$_{6}$, it could not be applied directly 
to Ba$_{0.9}$Sr$_{0.1}$CuSi$_{2}$O$_{6}$ because large oxygen bubbles from 
the reduction of CuO disconnected the two rods and terminated the crystal 
growth. In this sense, obtaining several gram-sized single crystals of 
Ba$_{0.9}$Sr$_{0.1}$CuSi$_{2}$O$_{6}$ constituted a significant achievement, 
made possible by using one single-crystalline seed in a mixed argon/oxygen 
atmosphere to reduce the sizes of the disruptive oxygen bubbles. 

\subsection{Inelastic neutron scattering measurements}

INS measurements at zero applied magnetic field were performed on the 
triple-axis spectrometers TASP and on the multiplexing triple-axis 
spectrometer CAMEA \cite{Groitl2016,Lass2021}, both at the SINQ neutron 
source \cite{Blau2009} at the Paul Scherrer Institute. INS probes the 
dynamical magnetic structure factor, $S({\vec Q},\omega)$, which is the 
spatial and temporal Fourier transform of the spin-spin correlation function, 
allowing direct access to the magnetic excitation spectrum as a function of 
the energy ($\hbar \omega$) and momentum ($\vec{Q}$) transfer of the scattered 
neutrons \cite{Furrer2009}. On TASP, a temperature of 1.5~K was used and the final 
neutron energy was fixed to 3.5~meV. The instrumental parameters of a previous 
study of BaCuSi$_{2}$O$_{6}$ \cite{Allenspach2020} were selected to optimize 
the energy resolution. On CAMEA, a temperature of 1.5~K was used and 
sequential measurements at fixed incident neutron energies of 7.5 and 9~meV 
were combined to obtain a map of the spectrum over multiple Brillouin zones 
and over an energy range from 2.5 to 5.5~meV. For both measurements, the 
rod-shaped single crystal was aligned with ($h$ 0 $l$) in the scattering 
plane [notation from the crystallographic unit cell, Fig.~\ref{fig1}(c)] 
\cite{Puphal2016}.

The CAMEA data shown in Figs.~\ref{fig1}(c)-\ref{fig1}(f) were extracted from 
the raw measured intensities using the software MJOLNIR \cite{Lass2020}, 
which corrects these intensities using measurements on a standard vanadium 
sample. The two datasets shown in Fig.~\ref{fig2}(a) differed by sample masses 
and counting times, which for the Ba$_{0.9}$Sr$_{0.1}$CuSi$_{2}$O$_{6}$ sample were 
measured for approximately 7.5 minutes per point, while the BaCuSi$_{2}$O$_{6}$ 
data were measured for approximately 15 minutes per point. The comparison was 
made by scaling the BaCuSi$_{2}$O$_{6}$ data so that the integrated intensity of 
all three modes matched the integrated intensity of the single mode observed 
in Ba$_{0.9}$Sr$_{0.1}$CuSi$_{2}$O$_{6}$.

\subsection{High-field measurements}

\noindent
{\bf{Magnetization.}}
Magnetization measurements up to 50~T were carried out in a capacitor-driven 
short-pulse magnet at the NHMFL Pulsed Field Facility in Los Alamos using an 
extraction technique \cite{Detwiler2000}. These measurements are subject to a 
strong reduction of the sample temperature in the changing magnetic field, 
which was calibrated by explicit MCE measurements.\\

\noindent
{\bf{Specific heat.}} The specific heat was measured with a heat-pulse 
technique inside a motor-generator-driven long-pulse magnet at the 
International MegaGauss Science Laboratory of the University of 
Tokyo \cite{Imajo2020}. The experimental data points were collected during 
the flat-topped portion of the magnetic-field profiles created by feedback loop 
control \cite{Kohama2015,Imajo2020}, at fields up to 37.8 T and temperatures 
down to 1 K.\\

\noindent
{\bf{Magnetic torque.}}
Measurements of the magnetic torque were conducted in a 31~T resistive magnet 
at the NHMFL DC Field Facility in Tallahassee using a silicon-membrane 
cantilever manufactured by NanoSensors$^{\rm TM}$ and a $^{3}$He cryostat to 
reach temperatures down to 0.3~K. Torque measurements were made during 
constant-temperature field sweeps across the phase boundary. The temperature 
was fixed by maintaining a constant $^3$He pressure in the sample space; 
two field-calibrated sensors, one located close to the cantilever and the 
other at the sample holder, were used for a precise in situ monitoring of the 
temperature. The angle between the field and the crystallographic axes of the 
sample was estimated by rotation tests to be 10.5$^\circ$. \\

\noindent
{\bf{Magnetocaloric effect (MCE).}}
MCE measurements were performed in vacuum at temperatures down to 1 K 
at the International MegaGauss Science Laboratory of the University of 
Tokyo \cite{Kihara2013}, and down to 0.4 K at the NHMFL in Los Alamos. In 
both experiments, the temperature was recorded during magnetic-field pulses 
up to 50 T using the response of a thin-film AuGe thermometer sputtered onto 
the sample surfaces. Under adiabatic vacuum conditions, the lines defined by 
$T(H)$ may be understood as isentropes. The sign of the temperature change is 
given by the Maxwell relation $(\partial Q/\partial H)/T = - (\partial M/
\partial T)|_H$; the changes of sign in the gradient of the isentropes 
correspond to the maxima or minima in $M(T)$ that occur at the ordering 
transitions.\\

\noindent
{\bf{Nuclear Magnetic Resonance (NMR).}}
$^{29}$Si NMR measurements were performed in the field range 22-26~T using a 
resistive magnet at LNCMI. A 13~mg single-crystal sample (5$\times$1$\times$0.5 
mm$^3$) was placed in the mixing chamber of a $^3$He-$^4$He dilution 
refrigerator with its $c$-axis oriented parallel to the applied field, 
${\vec H}$. NMR spectra were taken by the spin-echo pulse sequence. In the 
quantum disordered phase, the spectra are relatively narrow and were covered 
fully in a single-frequency recording; inside the magnetically ordered phase, 
the spectra are broader and their reconstruction required several recordings 
taken at equally spaced frequency intervals, by summation of the individual 
Fourier transforms. The frequency ($\nu$) of the $^{29}$Si spectra was measured 
relative to the reference $\nu_0/(\mu_0 H) =\,^{29}\gamma = 8.4577$~MHz/T. The 
magnitude of the applied field was calibrated using a metallic $^{27}$Al 
reference placed in the same coil as the sample. The same $^{27}$Al reference 
was also used to obtain a relative temperature calibration at temperatures 
below 1 K from the Boltzmann-factor dependence of the NMR signal intensity; 
this was then related to the values measured directly by the field-calibrated 
RuO$_2$ temperature sensor at all higher temperatures, where stable and 
field-independent $T$ values were ensured by fixing the pressure of the He bath.

\subsection{Quantum Monte Carlo}

We performed QMC simulations by applying the stochastic series expansion 
(SSE) algorithm \cite{Syljuasen2002} on 3D lattices of $L$$\times$$L\times$$L/2$
sites for $L = 8$, 10, 12, \ldots, 32. To reduce the computational cost we 
simulated an effective Hamiltonian of hard-core bosons \cite{Mila1998}, 
where each site represents a dimer and each site boson a dimer triplon. The 
dispersion relation of these bosonic quasiparticles was chosen to match the 
triplon band determined by INS.
 
For each applied field, the critical temperature, $T_c$, was extracted from 
the finite-size scaling of the superfluid density, $\rho_{\rm sf}$, of the 
hard-core bosons (equivalent to the spin stiffness of an ordered magnet). 
The quantity $L^{d + z - 2} \rho_{\rm sf}(L,T)$ has scaling dimension zero, 
where the dimensionality $d = 3$ and the dynamical exponent $z = 0$ for a 
finite-temperature transition, and hence the $L \rho_{\rm sf} (L,T)$ curves 
for all $L$ cross at the same point, which is $T = T_c$ \cite{Sandvik1998}.
We verified that the values of $T_c(H)$, and the associated uncertainties 
$\delta T_c(H)$, determined in this way were fully consistent with the 
critical exponent, $\nu = 0.672$, expected for the correlation length 
\cite{Allenspach2020}. 

\section{Magnetic excitations at zero field}
\label{sins}

The first step in characterizing the magnetic excitations of 
Ba$_{0.9}$Sr$_{0.1}$CuSi$_{2}$O$_{6}$ was to succeed in growing gram-sized 
single crystals by the technique described in Sec.~\ref{smm}A. 
Figures~\ref{fig1}(c) and \ref{fig1}(d) show the experimental neutron 
spectrum measured on CAMEA over multiple Brillouin zones for two high-symmetry 
directions, ($Q_h$ 0 4) and (0 0 $Q_l$), in reciprocal space. The spectrum along
($Q_h$ 0 4) displays one triplon mode with a periodicity of 2 in $Q_h$ and the 
spin gap (band minimum) at even values of $Q_h$ [Fig.~\ref{fig1}(c)]; this 
single mode has a periodicity of 4 in $Q_l$ and the spin gap appears at $Q_l
 = 2 + 4 n$ [Fig.~\ref{fig1}(d)], where $n$ is an integer. In our model of 
the spectrum, shown in Figs.~\ref{fig1}(e) and \ref{fig1}(f), one may state 
at the qualitative level that the location of the band minimum in $Q_h$ is a 
consequence of an effectively ferromagnetic (FM) intra-bilayer interaction 
parameter [$J^{\prime}$ in Fig.~\ref{fig1}(b)], as proposed \cite{Mazurenko2014} 
and observed \cite{Allenspach2020} in BaCuSi$_{2}$O$_{6}$. Similarly, the 
location of the band minimum in $Q_l$ is a consequence of a FM inter-bilayer 
interaction parameter, $J^{\prime\prime}$, which is the magnetic coupling between 
the top ions of dimers in one bilayer and the bottom ions of the dimers in the 
next [Fig.~\ref{fig1}(b)]. We remind the reader that the FM intra-bilayer 
correlations ensure that the inter-bilayer interactions are not frustrated 
for either sign of this latter parameter, excluding \cite{Mazurenko2014} the 
``dimensional reduction'' scenario mentioned above. The leading qualitative 
feature of the intensities is the modulation visible in Figs.~\ref{fig1}(c) 
and \ref{fig1}(e), which is a consequence of the fact that the resolution 
ellipsoid on CAMEA causes intensity to be spread over a narrower (focusing) 
or wider (de-focusing) range depending on the value of $\vec{Q}$.

Turning to a quantitative analysis of the low-temperature spectrum, 
Fig.~\ref{fig2}(a) shows the scattered intensity as a function of energy 
transfer for Ba$_{0.9}$Sr$_{0.1}$CuSi$_{2}$O$_{6}$ (blue symbols), and also 
the comparison with BaCuSi$_{2}$O$_{6}$ (red) \cite{Allenspach2020}, which we 
discuss in Sec.~\ref{sins}B; both data sets were measured on TASP at ${\vec Q}
 = (0 \; 0 \; 4)$ using similar experimental conditions. The solid lines are 
Gaussian fits used to extract the location, linewidth, and intensity of the 
triplon modes. The results of applying the same procedure at all ${\vec Q}$ 
points along the same two high-symmetry directions as in 
Figs.~\ref{fig1}(c)-\ref{fig1}(f) are shown as the solid 
symbols in Figs.~\ref{fig2}(b)-\ref{fig2}(e), where the locations and 
intensities are compared to the analogous results obtained from CAMEA. 
The triplon dispersion relations measured on the two spectrometers are 
in excellent agreement throughout the Brillouin zone [Figs.~\ref{fig2}(b) 
and \ref{fig2}(c)] and the measured intensities are fully compatible 
[Figs.~\ref{fig2}(d) and \ref{fig2}(e)].

\subsection{Spin Hamiltonian}

We stress that our observation of one triplon mode per minimal magnetic 
unit cell means that the Ba$_{0.9}$Sr$_{0.1}$CuSi$_{2}$O$_{6}$ structure contains 
only one type of bilayer, in contrast to the complicated situation in 
BaCuSi$_2$O$_6$ \cite{Allenspach2020}. An accurate determination of the spin 
Hamiltonian is crucial to verify the quasi-2D characteristics of the material 
and for a detailed investigation of the resulting critical behavior in an 
applied magnetic field. We assume that the minimal magnetic Hamiltonian of 
Ba$_{0.9}$Sr$_{0.1}$CuSi$_{2}$O$_{6}$ is that displayed in Fig.~\ref{fig1}(b), 
i.e.~that it contains only three interactions, all of purely Heisenberg type, 
to which, as above, we refer as intradimer ($J$), intra-bilayer ($J^{\prime}$), 
and inter-bilayer ($J^{\prime\prime}$). We note that $J^{\prime}$ is an effective 
inter-dimer interaction parameter that combines two pairs of ionic 
(Cu$^{2+}$-Cu$^{2+}$) interactions within and between the two layers of the 
bilayer structure \cite{Mazurenko2014,Allenspach2020}. 

By fitting the dispersion of the triplon mode shown in Figs.~\ref{fig2}(b) 
and \ref{fig2}(c) as detailed in App.~\ref{appins}, we deduce the optimal 
parameter set $J = 4.28(2)$~meV, $J^{\prime} = - 0.52(1)$ meV, and $J^{\prime\prime}
 = - 0.02(1)$~meV, where negative values denote FM interactions and the error 
bars represent a conservative estimate of the statistical uncertainties. 
Figures \ref{fig1}(e), \ref{fig1}(f), and \ref{fig2}(b)-\ref{fig2}(e) make 
clear that these parameters provide an excellent account of all the measured 
data, leaving no evidence of any requirement for further terms in the 
spin Hamiltonian. The INS parameters also serve as a benchmark for the 
magnetic interactions estimated using electronic structure calculations 
based on measurements of the atomic structure \cite{Puphal2016}, from which 
one may deduce (using the notation in Table II of that work) that $J_1 \equiv 
J$ is obtained to very high accuracy, $J_3 - J_5 \equiv J'$ is overestimated 
by approximately 30\%, $J_4 = 0$ is identified correctly, and $J_2 \equiv 
J^{\prime\prime}$ is surprisingly accurate in both size and sign, given its 
small value. On this point, we stress that the three terms we have deduced 
form a hierarchy of interactions whose strengths differ by at least an order 
of magnitude, and hence that each is responsible for physics at quite 
different energy scales. Specifically, from the standpoint of triplon 
excitations, Ba$_{0.9}$Sr$_{0.1}$CuSi$_{2}$O$_{6}$ is very much a quasi-2D 
magnetic material, with an inter-bilayer coupling 25 times smaller than 
the energy scale characterizing the intra-bilayer excitation processes.

\subsection{Comparison to BaCuSi$_2$O$_6$}

It is instructive to compare the magnetic interactions of the 
Ba$_{0.9}$Sr$_{0.1}$CuSi$_{2}$O$_{6}$ system with those of the parent compound. 
The transition to a weakly orthorhombic structure below 90~K \cite{Samulon2006,
Sparta2006,Stern2014} in BaCuSi$_2$O$_6$ results in the formation of three 
structurally inequivalent bilayers of dimers \cite{Rueegg2007,Sheptyakov2012}.
This led to the observation in INS experiments on BaCuSi$_{2}$O$_{6}$ of either 
three separate triplon modes or two separate peaks at ${\vec Q}$ vectors where 
the energies of the upper two modes are not distinguishable, which is the case 
in the data shown in Fig.~\ref{fig2}(a) for ${\vec Q} = (0 \; 0 \; 4)$. 

In Ba$_{0.9}$Sr$_{0.1}$CuSi$_{2}$O$_{6}$, the absence of this structural 
transition \cite{Puphal2016} means that the system remains tetragonal at 
low temperatures and only one bilayer type, hosting only one triplon mode, is 
expected. The neutron spectra presented in Figs.~\ref{fig1}(c)-\ref{fig1}(f) 
and \ref{fig2} confirm this situation. If the tetragonal interaction 
parameters are compared with the orthorhombic ones, $J = 4.28(2)$ is similar 
to $J_A = 4.275(2)$ meV, the weakest of the three intradimer couplings in 
BaCuSi$_{2}$O$_{6}$ \cite{Allenspach2020}, which explains the coincidence of 
the Ba$_{0.9}$Sr$_{0.1}$CuSi$_{2}$O$_{6}$ triplon shown in Fig.~\ref{fig2}(a) 
with the lower peak in the BaCuSi$_{2}$O$_{6}$ data. By contrast, $J^{\prime}
 = - 0.52(1)$ for the tetragonal structure is not close to $J^{\prime}_A
 = - 0.480(3)$~meV, but lies between the stronger interactions 
$J^{\prime}_B = - 0.497(8)$~meV and $J^{\prime}_C = - 0.57(1)$~meV found in the 
distorted structure. Finally, inter-bilayer interaction, $J^{\prime\prime} =
 - 0.02(1)$ meV, is only half the size of the corresponding value in 
BaCuSi$_{2}$O$_{6}$, which indicates the extreme sensitivity of these 
superexchange paths to the smallest structural alterations. However, while the 
bandwidth of the triplon mode visible in Figs.~\ref{fig1}(c), \ref{fig1}(e), 
and \ref{fig2}(b) is very similar to that in BaCuSi$_{2}$O$_{6}$, the bandwidth 
of the inter-bilayer dispersion [Figs.~\ref{fig1}(d), \ref{fig1}(f), and 
\ref{fig2}(c)] is much larger than the almost flat modes observed in 
BaCuSi$_{2}$O$_{6}$ \cite{Allenspach2020}. This is a consequence of the ABC 
bilayer stacking in BaCuSi$_{2}$O$_{6}$, by which the three different dimer 
types cause a dramatic reduction of the inter-bilayer triplon hopping 
probability.

In principle, valuable physical information is also contained in the line 
widths of the triplon modes observed in the two materials. Partial Sr 
substitution on the Ba sites should result in a degree of bond disorder, 
meaning a distribution in the values of the interaction parameters, which 
would act to broaden the triplon in Ba$_{0.9}$Sr$_{0.1}$CuSi$_{2}$O$_{6}$. 
Comparison with the \textit{A} triplon in BaCuSi$_{2}$O$_{6}$, which lies at 
the same energy in Fig.~\ref{fig2}(a), reveals that the Full Width at Half 
Maximum (FWHM) of the Gaussian characterizing the 10\%-substituted material is 
20\% higher at ${\vec Q} = (0 \; 0 \; 4)$ [0.32(2)~meV as opposed to 0.25(2) 
meV]. At ${\vec Q} = (1 \; 0 \; 4)$, however, the FWHM is essentially unchanged 
[0.30(2)~meV as opposed to 0.29(4)~meV, not shown]. A number of factors affect 
this comparison, including that the Sr-substituted crystal was subject to a 
grain reorientation during growth \cite{Puphal2019} and that the different 
shapes of the two crystals made it necessary to adjust the focusing slits 
for each case, affecting the experimental resolution. On the basis of these 
results, we are not able to discern a significant intrinsic broadening, due 
to bond disorder, from the extrinsic contributions to broadening, and thus 
we make only the qualitative statement that disorder effects on the magnetic 
properties arising from the weak structural disorder caused by Sr-substitution 
are small in Ba$_{0.9}$Sr$_{0.1}$CuSi$_{2}$O$_{6}$.

\begin{figure*}[t]
\noindent
\begin{centering}
\includegraphics[width=0.98\textwidth]{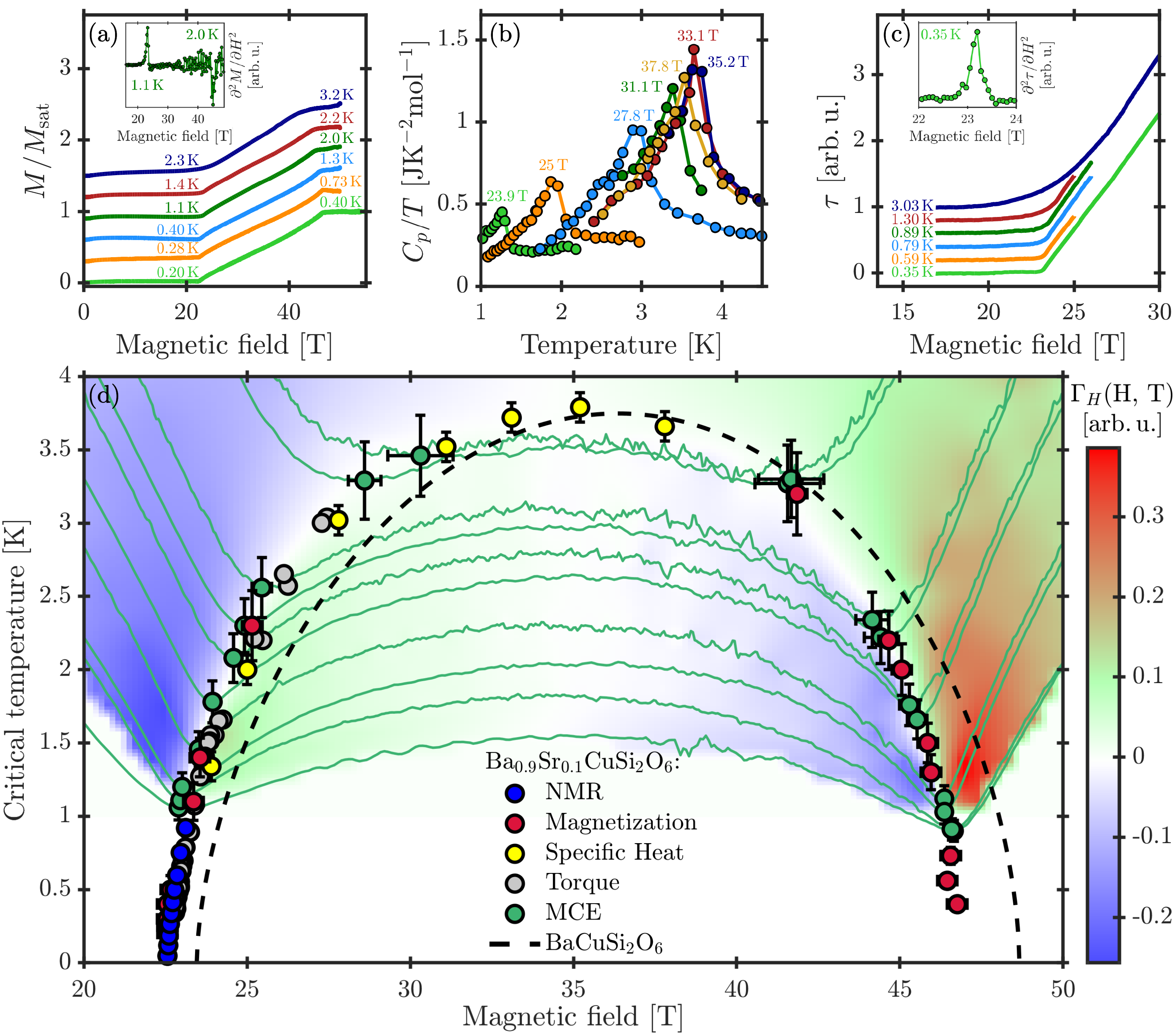} 
\end{centering}
\caption{{\bf Phase boundary of Ba$_{0.9}$Sr$_{0.1}$CuSi$_{2}$O$_{6}$.} 
Results from a wide range of thermodynamic measurements performed on 
single-crystalline Ba$_{0.9}$Sr$_{0.1}$CuSi$_{2}$O$_{6}$ at high magnetic fields, 
all with ${\hat H} \parallel {\hat c}$. (a) Magnetization, $M(H)$, measured in 
pulsed fields up to 50~T. Curves are shown with successive vertical offsets of 
0.3. Left and right labels are the calibrated temperatures obtained 
respectively for the lower and upper phase transitions after correction 
for the strong magnetocaloric effects (MCEs) acting during the field pulse. 
The inset shows the second derivative, ${\partial^2 M}/{\partial H}^2$, 
highlighting the locations of the ordering transitions. (b) Specific 
heat, $C_p(T)$, measured in pulsed fields up to 37.8~T and showing a 
$\lambda$-anomaly at the phase boundary. The maximum transition temperature, 
$T_{c}^{\rm max} \simeq 3.8$~K at 35.2 T, demarcates the top of the phase-boundary 
dome. (c) Magnetic torque, $\tau(H)$, measured in static fields around the 
onset of magnetic order. Curves above 0.35\,K are shown with vertical offsets 
of 0.2. The inset shows the second derivative, ${\partial^2 \tau}/{\partial 
H}^2$. (d) Field-temperature phase diagram of Ba$_{0.9}$Sr$_{0.1}$CuSi$_{2}$O$_{6}$ 
obtained by collecting all of our experimental data from magnetization, 
specific heat, magnetic torque, MCE, and NMR. Solid lines are lines of 
constant entropy obtained from our MCE measurements. Color contours represent 
the magnetic Gr\"uneisen parameter, $\Gamma_H = - (\partial M/\partial T)/C_p
 = - (\partial T/\partial H)_S/T$, which shows a sharp change of sign at the 
phase boundary. The black dashed line shows for comparison the phase boundary 
of BaCuSi$_{2}$O$_{6}$ \cite{Jaime2004}.}
\label{fig3}
\end{figure*}

\section{Field-temperature phase boundary} 
\label{spb}

The parameters we have determined for the gapped, zero-field ground state 
have essential consequences for the field-temperature phase diagram. The 
onset of field-induced magnetic order, when the field is strong enough 
to close the triplon gap, is generally referred to as a Bose-Einstein 
condensation of magnons; while the quadratic magnon dispersion at the 
transition ensures the Bose-Einstein universality class, the effective 
dimensionality of the system remains an open issue, as explained in 
Sec.~\ref{sintro}. Given the questions raised by the parent compound 
BaCuSi$_2$O$_6$, and the more general questions raised about possible 
fingerprints of BKT physics in coupled quasi-2D systems \cite{Furuya2016}, 
we have performed the most comprehensive study within our capabilities of 
the QPT and the surrounding quantum critical regime.

\subsection{High-field thermodynamic measurements}

For this we applied a combination of thermodynamic measurements up to the 
highest available pulsed and static magnetic fields (Sec.~\ref{smm}C) to map 
the field-temperature phase diagram of Ba$_{0.9}$Sr$_{0.1}$CuSi$_{2}$O$_{6}$. 
Magnetization data obtained in pulsed fields are shown in Fig.~\ref{fig3}(a). 
While zero magnetization indicates a spin gap, an abrupt jump and increasing 
$M(H)$ values are a signature of gap closure and magnetic polarization. The 
phase transitions both into and out of the ordered state are identified as 
sharp features in the second field derivative [inset, Fig.~\ref{fig3}(a)]. 
Although field pulses up to 50~T make it possible to reach saturation even at 
the lowest temperatures, the $M(H)$ curves taken in short-pulse magnets are 
adiabatic, rather than isothermal. BaCuSi$_2$O$_6$ and Ba$_{0.9}$Sr$_{0.1}$CuSi$
_{2}$O$_{6}$ have a strong MCE, meaning that the rapidly changing applied field 
drives a dramatic reduction of the sample temperature in the absence of heat 
transfer. We performed explicit MCE measurements (below) in order to obtain 
an accurate determination of the real measurement temperatures shown for 
both phase transitions in Fig.~\ref{fig3}(a).

We have measured the specific heat in pulsed fields up to 37.8~T and display 
the results in Fig.~\ref{fig3}(b). Because this field value is closer to the 
upper critical field of Ba$_{0.9}$Sr$_{0.1}$CuSi$_{2}$O$_{6}$ than to the lower, 
these measurements make it possible to access the top of the phase-boundary 
``dome'' shown in Fig.~\ref{fig3}(d). The $\lambda$-shaped anomalies in 
$C_p(T)$ are characteristic of second-order phase transitions and we used 
an equal-area method to estimate the critical temperatures for each magnetic 
field. The $\lambda$ shapes have also been associated directly with the 3D XY 
universality class in a number of systems showing magnon BEC \cite{Oosawa2001,
Jaime2004,Sebastian2005,Aczel2009a,Aczel2009b,Kohama2011,Weickert2019}, while 
below the transition $C_p(T)$ is consistent with the $T^3$ form expected from 
AF magnons in 3D. 

In a static external field, we have measured the magnetic torque up to 29 T, 
as shown in Fig.~\ref{fig3}(c). In these experiments the sample temperature is 
known exactly but, because a torque acts only when the sample is tilted away 
from its principal axes, the tilt angle provides a degree of uncertainty that 
affects the absolute field. By contrast, the relative values of the critical 
fields at each temperature are extremely accurate, allowing one to probe the 
shape of much of the left side of the phase-boundary dome. Again the second 
field-derivative [inset Fig.~\ref{fig3}(c)] gives the most accurate 
determination of the critical field, $H_c$, for each measurement temperature, 
$T$. Here we have corrected these fields to the NMR values of $H_c(T)$ (below) 
in the regime of temperature where the two methods overlap, thereby obtaining 
the phase-boundary points shown up to 3.03 K in Fig.~\ref{fig3}(d). 

The MCE, meaning the change in sample temperature due to an applied field, 
is related directly to the first temperature derivative of the magnetization 
(Sec.~\ref{smm}C). MCE measurements in pulsed fields up to 50 T were used to 
obtain the lines of constant entropy shown in Fig.~\ref{fig3}(d), whose local 
minima mark the location of the ordering transition in both field and 
temperature. We carried out two separate MCE experiments, at ISSP Tokyo 
and NHMFL Los Alamos, and used the latter data to deduce the true sample 
temperature for the magnetization results shown in Fig.~\ref{fig3}(a). The MCE 
shows a discernible asymmetry [Fig.~\ref{fig3}(d)], in that the phase-boundary 
temperature for the same isentrope is higher on the left side of the dome 
than on the right \cite{Nomura2020}, indicating that the isothermal entropy, 
$S(H)$, is higher on the right than on the left. This result implies a larger 
specific heat on the right side of the dome and has been explained as a magnon 
mass anisotropy \cite{Kohama2011} that is connected to the ratio of the upper 
and lower critical fields. In Fig.~\ref{fig3}(d) we show also the magnetic 
Gr\"uneisen parameter, whose sign change becomes increasingly abrupt as the 
temperature is lowered, thereby locating the phase boundary with increasing 
accuracy. 

We also performed NMR measurements at very low temperatures to determine 
the phase boundary close to the lower critical field, but defer a detailed 
discussion of these experiments to the next subsection. Figure~\ref{fig3}(d) 
compiles all of our experimental data to display a comprehensive picture of 
the dome-shaped phase boundary of the magnetically ordered (condensate) phase.
The dome is rather symmetrical in shape between a lower critical field 
$\mu_0 H_{c1} \simeq 22.5$ T and an upper critical field $\mu_0 H_{c2} \simeq 
46.5$ T, with a maximum height of $T_{c}^{\rm max} \simeq 3.8$~K around 35 T. 
In Fig.~\ref{fig3}(d) we show also the phase-boundary dome determined
\cite{Jaime2004} for the parent compound, BaCuSi$_{2}$O$_{6}$, which is 
altered only slightly by the Sr substitution. Specifically, the small 
reduction of $H_{c1}$ is a consequence of the smaller gap in 
Ba$_{0.9}$Sr$_{0.1}$CuSi$_{2}$O$_{6}$, which arises due to the combination 
of increased bandwidth (controlled by $J^{\prime}$) and rather constant 
band center (controlled by $J$). The reduction of the upper critical field, 
$H_{c2}$, is a consequence of the fact that the stronger intra-dimer coupling 
constants of the structurally inequivalent bilayers in BaCuSi$_{2}$O$_{6}$ are 
absent in Ba$_{0.9}$Sr$_{0.1}$CuSi$_{2}$O$_{6}$. 

\begin{figure*}[t]
\noindent
\begin{centering}
\includegraphics[width=0.98\textwidth]{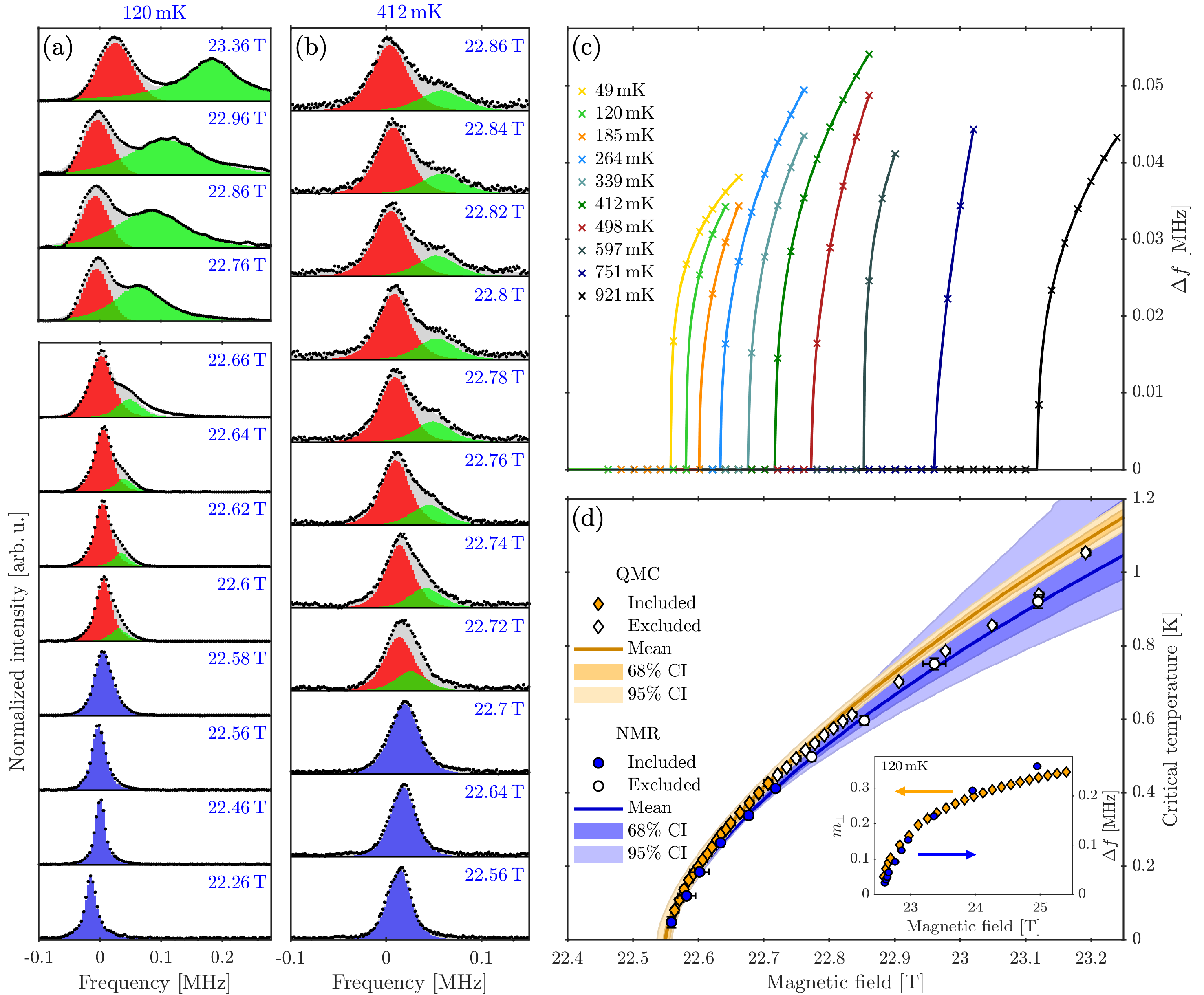} 
\end{centering}
\caption{{\bf NMR measurements in the critical regime.} (a)-(b) $^{29}$Si NMR 
spectra measured for a range of field values around the corresponding value 
of $H_c(T)$ at 120~mK (a) and 412~mK (b). The spectra show a splitting from 
a single, sharp peak (blue) in the quantum disordered phase at $H < H_{c}(T)$ 
to two generally broader peaks (red and green) in the magnetically ordered 
phase at $H > H_{c}(T)$. The measured intensities were modelled as the sum 
(grey) of the single or double peak(s) and a weak, constant contribution. 
The double peaks were modelled by a simultaneous fitting procedure in all 
panels other than the highest four fields shown at 120 mK (a), where 
individual peak fits were used. (c) Splitting of peaks in the NMR spectra. 
Solid lines show splitting functions deduced from simultaneous fits to all 
spectra measured at the same temperature, and crosses mark the field values 
at which data were taken. (d) Low-field phase boundary of 
Ba$_{0.9}$Sr$_{0.1}$CuSi$_2$O$_{6}$ extracted from the NMR peak-splitting data 
(blue and white circles). Orange and white squares show the phase boundary 
obtained from QMC simulations. Blue and orange lines show the optimal fit 
to the phase boundary based on each set of points, obtained from the mean 
of the posterior distribution determined by a Bayesian inference procedure 
based on Eq.~\eqref{eq:phase_boundary}. The blue and orange shaded areas 
represent the estimated uncertainties in the form of the 68\% (dark) and 
95\% (light shading) credible intervals (CIs). NMR and QMC data points 
marked by white circles and squares are those excluded from the statistical 
analysis of the quantum critical properties. Inset: NMR peak-splitting 
measured up to 25 T at $T = 120$ mK (blue circles) juxtaposed with the 
magnetic order parameter, $m_\perp (T)$, obtained from QMC simulations.}
\label{fig4}
\end{figure*}

\subsection{Nuclear Magnetic Resonance}

For a detailed investigation of the quantum critical regime, we performed 
NMR measurements at high magnetic fields and very low temperatures using the 
facilities at the LNCMI (Sec.~\ref{smm}C). In the magnetically ordered phase 
above the critical field, which at any temperature we denote by $H_c(T)$, the 
presence of a staggered magnetization, $m_\perp(T)$, transverse to the 
direction of the external field (which is applied along the ${\hat c}$-axis) 
lifts the degeneracy between sites that were equivalent in the gapped dimer 
phase \cite{Berthier2017,Arcon2017}. Their NMR spectra therefore show a 
splitting into two separate peaks. Figure \ref{fig4}(a) shows $^{29}$Si NMR 
spectra obtained using a 13~mg single crystal of Ba$_{0.9}$Sr$_{0.1}$CuSi$
_{2}$O$_{6}$ at $T = 120$ mK over a range of static magnetic fields crossing 
the value of $H_{c}$ for this temperature. Whereas in BaCuSi$_{2}$O$_{6}$ it 
was possible by $^{29}$Si NMR to detect only a broad distribution due 
to the presence of structurally different bilayer types \cite{Horvatic2005,
Kraemer2007,Stern2014}, whose local spin polarization could be determined 
quantitatively only by $^{63}$Cu NMR \cite{Kraemer2013}, in the $^{29}$Si 
spectra of the Sr-substituted variant we observe a second peak that splits 
from the first with increasing applied field, becoming a separate and clearly 
resolved entity in the upper panels of Fig.~\ref{fig4}(a). 

Because of the low temperatures of these measurements, the field value at 
which the NMR peak-splitting appears constitutes our most accurate estimate 
of the critical field, $H_{c}(T)$, for each temperature, $T$, in the regime 
most likely to approach quantum critical scaling. The splitting of the peaks 
in frequency is directly proportional to the magnetic order parameter 
\cite{Berthier2017}, allowing us to obtain the field-induced evolution of 
the ordered moment up to a single, non-universal constant of proportionality, 
which is the hyperfine coupling constant of the NMR experiment. To determine 
the field-dependence of the peak-splitting, and hence of the order parameter, 
we measured NMR spectra up to fields significantly higher than $H_{c}(T)$ for 
one example temperature, 120 mK [Fig.~\ref{fig4}(a)], and the results are shown 
in the inset of Fig.~\ref{fig4}(d). To determine the phase boundary for a study 
of its quantum critical properties, we have measured NMR spectra for a number 
of fixed temperatures between 49 and 921~mK, over a range of fields close to 
$H_c(T)$ in each case, and a further set of spectra is shown for $T = 412$ mK 
in Fig.~\ref{fig4}(b). 

To optimize our extraction of the peak-splitting from the spectral data, which 
contain multiple additional features and complications, we exploit its relation 
to the order parameter to model it as a power-law function of the field 
difference from the phase transition, $H - H_{c}(T)$, and perform a simultaneous
fit of one peak (blue) or two split peaks [red and green in Figs.~\ref{fig4}(a) 
and \ref{fig4}(b)] to all NMR spectra measured at the same $T$. Details of this 
procedure, which is valid over a field-temperature regime near the phase 
boundary, are provided in App.~\ref{appnmrpb}. Only at fields well above 
$H_c(T)$ was it possible to obtain reliable fits to the two clearly separated 
spectral peaks of each individual spectrum, and examples of these fits are 
shown in the upper panels of Fig.~\ref{fig4}(a). 

The result of the simultaneous fit procedure is the set of fitting functions 
shown by the lines in Fig.~\ref{fig4}(c), where the crosses represent the 
fields at which spectra were collected, and these lines yield a set of 
estimated critical points, $H_{c,j}$, with uncertainties $\delta H_{c,j}$. The 
corresponding temperature values, with their uncertainties, were determined 
relative to a field-calibrated RuO$_2$ thermometer (Sec.~\ref{smm}C). The 
data pairs $(H_{c,j}, \delta H_{c,j})$ and $(T_{c,j}, \delta T_{c,j})$ together 
constitute the NMR phase boundary near the quantum critical point, which we 
show as the blue and open circles in Fig.~\ref{fig4}(d). The blue line and 
shading represent the results of our statistical analysis of the quantum 
critical properties of the NMR phase boundary, which we discuss below. 

To assist with the interpretation of the measured phase boundary, we 
performed large-scale QMC calculations of a model with only one type 
of bilayer, following the established procedure \cite{Allenspach2020} 
summarized in Sec.~\ref{smm}D. For this we mapped the spin Hamiltonian of 
Fig.~\ref{fig1}(b) to a hard-core-boson model \cite{Mila1998} for triplons 
with the low-energy triplon dispersion determined from our zero-field INS 
measurements. The QMC phase boundary in the regime close to the quantum 
critical point (QCP) is shown by the orange and open squares in 
Fig.~\ref{fig4}(b), while the corresponding lines and shading are the 
results of the statistical analysis we present next (Sec.~\ref{sce}). We 
remark that the precise agreement in location of the QCP obtained by using 
the zero-field parameters implies that magnetostriction effects are 
negligible, a result also found in BaCuSi$_{2}$O$_{6}$. We also performed 
QMC simulations to determine the order parameter at low temperatures for 
fields well inside the magnetically ordered regime, and in the inset of 
Fig.~\ref{fig4}(d) these are shown for a temperature corresponding to 
120 mK. By matching the field-induced ordered moment in Bohr magnetons to 
the peak-splitting measured in MHz over this field range, we obtain the 
hyperfine coupling constant as noted above. 

\section{Critical exponent}
\label{sce}

From the statistical mechanics of classical phase transitions, the 
field-temperature phase boundary should be described by a power-law form 
in the critical regime close to the QCP, and hence can be expressed as
\begin{equation}
T_c(H) = \alpha \big( H - H_{c1} \big)^{\phi}.
\label{eq:phase_boundary}
\end{equation}
Here $H_{c1}$ is the critical field of the QCP, i.e.~at zero temperature, 
$\alpha$ is a nonuniversal constant of proportionality and $\phi$ is the 
critical exponent, which is universal and should reflect only the 
dimensionality of space and the symmetry of the order parameter. In a true 
BEC one has $\phi = 2/d$ \cite{Nohadani2004,Kawashima2004,Sebastian2006,
Laflorencie2009}, where the numerator is the dynamical exponent for free 
bosons ($\omega \propto k^2$) and $d$ the dimensionality of space. The form 
of Eq.~\eqref{eq:phase_boundary} means that the parameters ($\alpha$, 
$H_{c1}$, $\phi$) required to model the experimental data are highly 
correlated; specifically, the best estimate for $\phi$ depends strongly on 
the value of $H_{c1}$. Further, the width of the quantum critical regime, the 
ranges of $T_c$ and $H - H_{c1}$ over which Eq.~\eqref{eq:phase_boundary} 
remains valid, is a non-universal quantity about which little specific 
information is available. Attempts to deal with both the correlation 
problem \cite{Sebastian2006} and the critical-regime problem \cite{Tanaka2001,
Oosawa2002} on the basis of limited available data have led in the past to 
significant confusion, which has been cleared up for certain 
models \cite{Nohadani2004}. In order to obtain an accurate estimate 
of $\phi$ and to quantify its uncertainty reliably, we have applied 
a thorough statistical analysis using Bayesian inference.

Bayesian inference is a general method for the statistical analysis of all 
forms of experimental data, and thus its applications can be found throughout 
physics \cite{Toussaint2011}. The quantity of interest is the posterior 
probability distribution, 
\begin{equation}
p (\bm{\theta} | \mathcal{D}) \propto p (\mathcal{D} | \bm{\theta}) p 
(\bm{\theta}),
\label{eq:BayesTheorem}
\end{equation}
of the model parameters, $\bm{\theta}$, conditional on the observed data, 
$\mathcal{D}$. In this expression of Bayes' theorem, $p (\mathcal{D} | 
\bm{\theta})$ denotes the likelihood of making the observations $\mathcal{D}$ 
for a given $\bm{\theta}$ and $p (\bm{\theta})$ is a prior probability 
distribution summarizing any information previously available about the 
model parameters. In the problem of describing the phase boundary, 
$\bm{\theta} = (\alpha, H_{c1}, \phi)$ is the set of parameters in the model 
of Eq.~\eqref{eq:phase_boundary} and $\mathcal{D} = \{H_{c,j}, T_{c,j}, \delta 
H_{c,j}, \delta T_{c,j}\}$ is the set of phase-boundary points with their 
associated uncertainties; clearly the total number of points ($j$) in the 
input information will have a leading role in determining the accuracy of 
the output. The posterior distribution, $p (\bm{\theta} | \mathcal{D})$, is 
a high-dimensional density function that contains all of the information 
available about the parameters in $\bm{\theta}$, including their uncertainties 
and (non-linear) correlations. To obtain individual parameter estimates and 
error bars, the posterior distribution is compressed in various ways. We use 
the mean of the marginal distribution of each parameter, which provides the 
prediction with the smallest error given the input data \cite{Toussaint2011}. 
To quantify these errors, we construct the highest posterior-density credible 
intervals (CIs), which are the shortest intervals containing a fixed fraction 
of the posterior density \cite{Gelman2004}, and show the commonly chosen 68\% 
and 95\% CIs.

Figure \ref{fig4}(d) shows the mean of the posterior distributions of the 
model function of Eq.~\eqref{eq:phase_boundary} as a blue line for the 
NMR dataset, $\mathcal{D}_{\rm NMR}$, and an orange line for our QMC data, 
$\mathcal{D}_{\rm QMC}$, while the shaded regions indicate the 68\% and 95\% 
CIs. Referring back to the issue of the width of the quantum critical regime, 
the two posterior distributions were constructed on the basis of the 6 NMR 
phase-boundary data points falling below the temperature $T_{{\rm max}}^{{\rm NMR}}
 = 412$ mK (blue circles) and the 17 QMC points below $T_{{\rm max}}^{{\rm QMC}} =
426$ mK (orange squares). The criterion by which these maximum values were 
determined is discussed below, and an extension of the fit to higher 
temperatures is shown for reference. 

\begin{figure*}[t]
\noindent
\begin{centering}
\includegraphics[width=\textwidth]{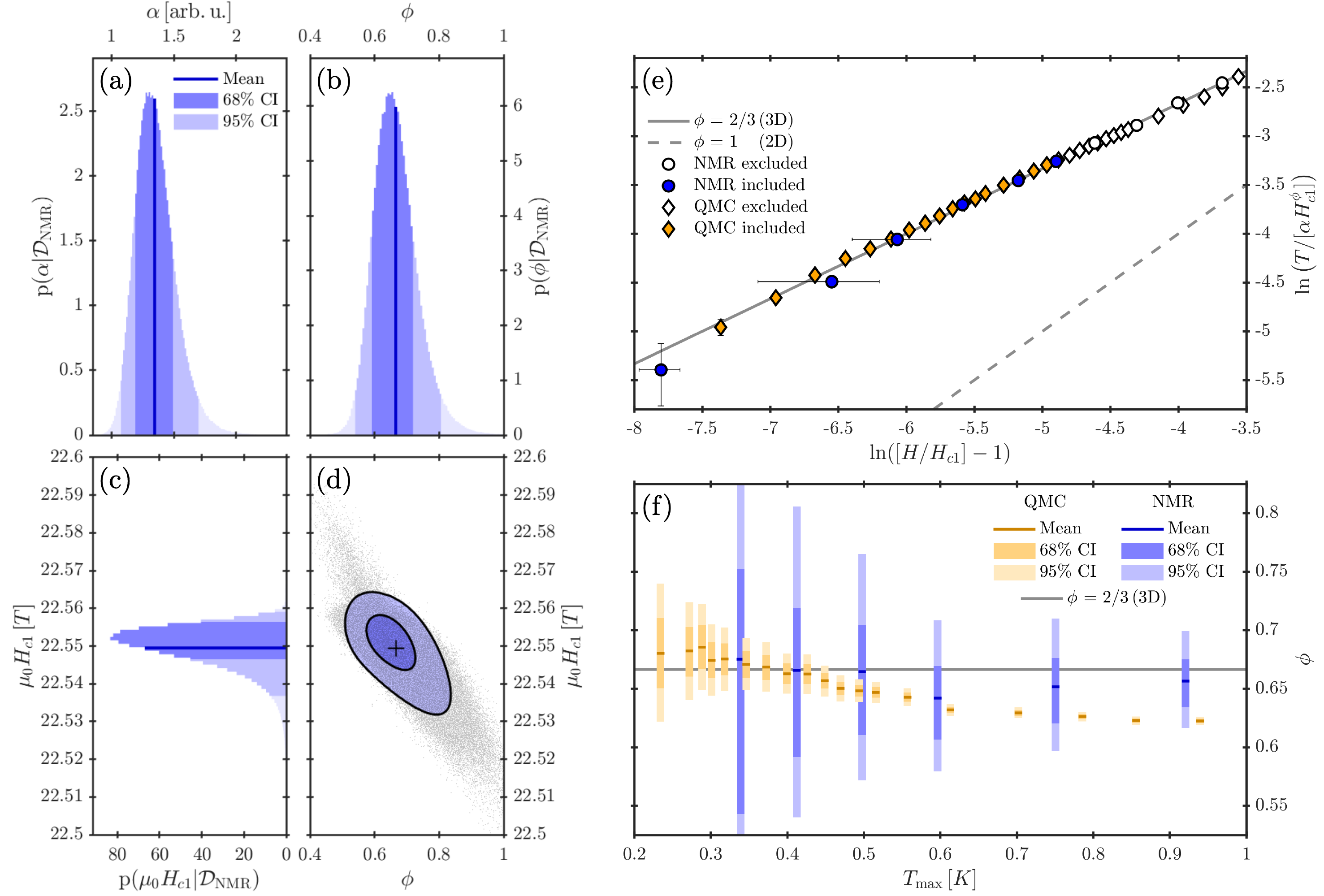} 
\end{centering}
\caption{{\bf Analysis of the critical exponent.} (a)-(c) Posterior probability 
distributions of the parameters in the model of Eq.~\eqref{eq:phase_boundary} 
for the field-temperature phase boundary. The input for the analysis is the 
set, $\mathcal{D}_{\rm NMR}$, of 6 blue NMR data points up to $T_{\rm max} = 
412$~mK in Fig.~\ref{fig4}(d). The marginal posterior distributions, 
$p(\theta_i|\mathcal{D})$, for each of the model parameters $\theta_i \equiv 
\alpha$ (a), $\phi$ (b), and $H_{c1}$ (c), are visualized as histograms of 
samples drawn from the posterior distribution together with their mean (blue 
lines), 68\%, and 95\% CI boundaries. (d) Posterior distribution shown by 
projection to $\phi$ and $H_{c1}$ (i.e.~integrating over $\alpha$), which 
illustrates the strong negative correlation between the two parameters. 
(e) Phase-boundary points of Fig.~\ref{fig4}(d) shown on logarithmic axes. 
The gradients of both sets of points correspond to the critical exponent, 
$\phi$. For comparison we indicate the cases $\phi = 1$ (dotted) and $\phi
 = 2/3$ (dashed lines). (f) Evolution of $\phi$ as a function of the maximal 
temperature, $T_{{\rm max}}$, up to which phase-boundary data points are included.
Horizontal lines show the mean and shaded boxes the 68\% and 95\% CIs of the 
marginal distribution of $\phi$ deduced by including only points below 
$T_{\rm max}$; the 95\% CIs for $T_{\rm max} = 264$ mK lie at 0.465 and 0.923. 
The NMR symbols were obtained by including 5, 6, ... 10 of the data points 
shown in Fig.~\ref{fig4}(d) and panel (e), the QMC symbols by including 
between 8 and 30 data points.}
\label{fig5}
\end{figure*}

To explain these results, Figs.~\ref{fig5}(a)-\ref{fig5}(c) show the marginal 
distributions, $p(\theta_i|\mathcal{D}_{\rm NMR})$, of each of the three model 
parameters in $\bm{\theta}$. These are normalized density distributions, 
obtained from Monte Carlo sampling of the posterior by integrating over the 
remaining two parameters in each case, which allow the full shape of each 
marginal to be investigated. These are clearly skewed, with the distributions 
of $\alpha$ [Fig.~\ref{fig5}(a)] and $\phi$ [Fig.~\ref{fig5}(b)] having tails 
extending towards larger values whereas that of $H_{c1}$ [Fig.~\ref{fig5}(c)] 
extends towards smaller values. Qualitatively, the origin of these distorted 
shapes lies in the interplay of the uncertainties in the field and temperature 
measurements, whose simultaneous consideration is a key strength of the 
Bayesian approach; details of its quantitative implementation are presented 
in App.~\ref{appbayes}. By contrast, a regression analysis such as a 
least-squares approach would yield biased estimates under these circumstances. 
Figure~\ref{fig5}(d) shows a projection of the posterior distribution as a pair 
plot depending on $\phi$ and $H_{c1}$, which highlights the strong negative 
correlation between these two parameters. In the absence of accurate knowledge 
of $H_{c1}$, the NMR phase boundary data of Fig.~\ref{fig4}(d) can often be 
described by decreasing $H_{c1}$ while simultaneously reducing the height of 
the curve, i.e.~increasing $\phi$, or vice versa. In view of this situation, 
one may consider the precise goal of the present analysis as being to localize 
the most probable ($H_{c1}$, $\phi$) pair accurately on the axes of 
Fig.~\ref{fig5}(d).

Using the mean and 68\% CI boundaries of the marginal posterior distributions,
from our NMR data for $T \leq T_{\rm max}^{\rm NMR}$ we obtain the model parameter 
estimates $\alpha = 1.35^{+0.15}_{-0.16}$ K, $\mu_0 H_{c1} = 22.550^{+0.007}_{-0.003}$ 
T, and $\phi = 0.666^{+0.053}_{-0.074}$. The non-universal parameter $\alpha$ has 
no influence on the critical properties. By comparing $H_{c1}$, for a field 
applied along the ${\hat c}$-axis, with the value of the spin gap extracted 
from INS, $\Delta = 3.00(1)$~meV, we deduce a $g$-factor of $g_{\parallel c}
 = 2.298^{+0.008}_{-0.008}$. This result is in agreement with both the value 
2.32(2) determined for the Sr-doped material \cite{Puphal2016} and the value 
$g_{\parallel c} = 2.306(3)$ obtained for the parent compound \cite{Zvyagin2006}. 
Clearly the mean value of $\phi$ we obtain agrees spectacularly well with 
the 3D exponent of 2/3, although we stress that the CIs are broad (as one 
may expect from the number of data points in the analysis), and to interpret 
these results we turn to QMC modelling.

Because our QMC simulations were performed with an effective quasiparticle 
Hamiltonian matching the INS dispersion, $H_{c1}$ is fixed by the lower band 
edge (using the same $g$-factor) and the set of points $\{H_{c,j}\}$ has no 
errors ($\{\delta H_{c,j}\} = 0$), which simplifies the statistical analysis 
(App.~\ref{appbayes}). We deduce the values $\alpha = {\rm 1.46}^{+0.03}_{-0.03}$ 
K and $\phi = 0.663^{+0.008}_{-0.009}$, based on our data below $T_{{\rm max}}
^{{\rm QMC}} = 426$ mK. Figure \ref{fig5}(e) shows the putative critical regime 
on logarithmic axes, a format in which the gradient of each line is the 
critical exponent, $\phi$, and in which both our NMR and QMC results are 
indistinguishable from the line $\phi = 2/3$ expected of a 3D BEC. 

Returning to the width of the critical regime, this is a non-universal quantity 
in studies of critical behavior and hence is an unknown in any fitting 
procedure. Under these circumstances, the windowing fit was introduced 
\cite{Nohadani2004} to discern the trend of the data towards a critical 
value at the QCP when the width of the critical regime may be arbitrarily 
small, and has been applied \cite{Sebastian2006,Allenspach2020} for the 
analysis of BaCuSi$_2$O$_6$ data. To investigate this issue, we performed an 
effective windowing analysis by applying the Bayesian inference procedure using 
10, 9, 8, 7, 6, and 5 NMR data points, and the results for the fitted exponents 
are shown as the blue symbols in Fig.~\ref{fig5}(f). The horizontal lines show 
the mean and the heights of the shaded boxes show the 68\% and 95\% CIs of the 
marginal posterior distribution of $\phi$ for each case. We observe that the 
estimated mean value of $\phi$ is almost exactly 2/3 for the three narrowest 
windows, but then dips on inclusion of the 8th data point and recovers 
slightly towards 2/3 when the 9th and 10th points are considered. However, 
by a criterion that the mean $\phi$ should include 2/3 within the 68\% CI, 
all of these windows are consistent with 3D critical scaling.

For further insight as to whether it is realistic to ascribe this behavior 
in its entirety to quantum critical physics, we perform the same type of 
analysis using our QMC data. While QMC studies of 3D models have revealed 
quantum critical regimes with relative widths on the order of 10-20\% of 
the control parameter \cite{Nohadani2004,Qin2015}, simulations of quasi-1D 
models have struggled to reach such a regime. Our QMC results, which have 
better data coverage and much narrower CIs, can be expected to reflect the 
physics of quasi-2D bosons at the energy scales of their interlayer coupling. 
The data in Figs.~\ref{fig5}(f) show a systematic and near-monotonic evolution 
best described as the estimated $\phi$ converging to 2/3 from below as the 
number of data points is reduced. If one defines $T_{\rm max}$ as the highest 
temperature at which $\phi$ has converged to 2/3 within the 68\% CI, one 
obtains $T_{\rm max}^{{\rm QMC}} = 426$ mK, which corresponds to the use of 17 
phase-boundary points. On this basis we deduce that effects beyond quantum 
critical fluctuations become discernible above approximately 0.4 K, and 
hence that this is an appropriate width to adopt for the scaling regime. 

This criterion terminates our NMR dataset at $T_{\rm max}^{\rm NMR} = 412$ mK, 
corresponding to the 6 phase-boundary points used for the analysis shown in 
Figs.~\ref{fig4}(d) and \ref{fig5}(a)-\ref{fig5}(d). However, we note that 
the criterion is far from exact, and in the present case an almost identical 
estimate with lower uncertainties, $\phi = 0.665^{+0.040}_{-0.054}$, is obtained 
by including the 7th data point. The evolution of the estimated $\phi$ on 
inclusion of the 9th and 10th data points indicates both the onset of complex 
and manifestly non-monotonic crossover behavior beyond the critical regime and 
the accuracy of the analysis beyond features visible in Figs.~\ref{fig4}(d) or 
Figs.~\ref{fig5}(e). The discrepancy between the forms of behavior exhibited 
by our NMR and QMC data could lie in interaction terms neglected in our 
minimal magnetic Hamiltonian, or in possible disorder effects arising due 
to Sr substitution, which we discuss further in Sec.~\ref{sd}. However, we 
stress that these differences in critical behavior are extremely subtle, and 
are certainly not discernible in the NMR and QMC phase boundaries shown in 
Fig.~\ref{fig4}(d).

To summarize, our NMR measurements and statistical analysis, backed up by 
our QMC simulations, present a quite unambiguous demonstration of 3D critical 
physics in a material with a strongly quasi-2D structure and magnetic 
properties. With reference to the prediction that unconventional magnetic 
ordering bearing the hallmarks of BKT physics should appear uniquely in 3D 
coupled quasi-2D systems \cite{Furuya2016}, our results [Figs.~\ref{fig3}(c) 
and \ref{fig4}(d)] demonstrate that the field-induced onset and evolution of 
the ordered phase are quite conventional. From this we conclude that the 
inter-bilayer coupling in Ba$_{0.9}$Sr$_{0.1}$CuSi$_{2}$O$_{6}$ is still 
significantly too large to access BKT physics, despite being only 1/100 
of the magnon bandwidth (1/25 of the intra-bilayer interaction, a value 
that by any other measure appears worthy of the designation ``quasi-2D''). 
Nevertheless, the restoration of observable 3D quantum criticality from an 
apparently minor chemical substitution constitutes a somewhat remarkable 
consequence of structural ``disorder.'' 

\section{Discussion}
\label{sd}

We have found that the interaction parameters of 
Ba$_{0.9}$Sr$_{0.1}$CuSi$_{2}$O$_{6}$ determined by INS at zero field, 
$J = 4.28(2)$ meV, $J^{\prime} = - 0.52(1)$ meV, and $J^{\prime\prime} = - 0.02(1)$ 
meV, provide an excellent description not only of the triplon spectrum 
(Figs.~\ref{fig1} and \ref{fig2}) but also of the critical properties at 
the field-induced QPT [Figs.~\ref{fig4}(d) and \ref{fig5}]. It is notable 
that these interaction parameters are not changed, within the error bars 
of our zero- and high-field experiments, by applied fields in excess of 20 T. 
More noteworthy still is the hierarchy of energy scales contained within the 
minimal spin Hamiltonian, which appear to define a quasi-2D material. The 
zero-field spectrum measured in Figs.~\ref{fig1} and \ref{fig2} probes largely 
the intra-bilayer energy scale, which gives the triplons a bandwidth of 2 meV. 
However, it is clear in Fig.~\ref{fig5}(f) that the inter-bilayer energy of 
fractions of a Kelvin is the scale on which the 3D physics emerges. 

We remark again that the FM effective intra-bilayer interaction removes any 
frustration of the inter-bilayer interactions, reinforcing the expectation of 
a simple scaling form around the QCP. In a windowing analysis of the critical 
scaling in BaCuSi$_{2}$O$_{6}$, an abrupt change in behavior towards an 
exponent $\phi = 1$ was reported below 0.75 K, which was interpreted as a 
frustration-induced dimensional reduction \cite{Sebastian2006}. Subsequent 
investigation has revealed the absence of frustration \cite{Mazurenko2014}, the 
presence of structurally inequivalent bilayers \cite{Rueegg2007,Sheptyakov2012},
and the problem of the strong negative correlation between $\phi$ and $H_{c1}$ 
[shown clearly in our Fig.~\ref{fig5}(d)]. The experimental confirmation of 
the absence of frustration was accompanied by detailed QMC simulations of the 
quantum critical regime, from which it was possible to conclude that the true 
behavior of the system includes an intermediate field-temperature regime 
of non-universal scaling governed by an anomalous exponent whose value 
turned out to be $\phi \gtrsim 0.72$ over much of the effective scaling 
range \cite{Allenspach2020}. This result raises the question of whether our 
present analysis, which obviously relies on a limited number of experimental 
data points, can be regarded as sufficiently precise to claim that 
Ba$_{0.9}$Sr$_{0.1}$CuSi$_{2}$O$_{6}$ shows true 3D scaling. For this we appeal 
first to the fact we demonstrated by INS, that the Sr-substituted material 
has only one bilayer type and one species of triplon (Figs.~\ref{fig1} and 
\ref{fig2}), whereas all the complications leading to anomalous scaling in 
BaCuSi$_{2}$O$_{6}$ result from a new effective energy scale dictated by the 
separation of the inequivalent triplon modes. Thus in the Sr-substituted 
case one expects both a wider critical regime and a lack of alternatives 
other than the universal 2D or 3D forms, and our results are certainly not 
consistent with 2D scaling ($\phi = 1$). Second, we note from the error bars in 
the QMC data for the layered model of BaCuSi$_{2}$O$_{6}$ \cite{Allenspach2020} 
that the anomalous effective exponent is in no way compatible with 2/3; by 
contrast, our NMR data for Ba$_{0.9}$Sr$_{0.1}$CuSi$_{2}$O$_{6}$ favor strongly 
an exponent of 2/3 in Fig.~\ref{fig5}(f) rather than any anomalous value, as 
do our QMC simulations for the model with only one type of bilayer.

The crux of our analysis is the materials-synthesis result that Sr substitution
improves the regularity of the BaCuSi$_{2}$O$_{6}$ structure, maintaining 
tetragonal symmetry and eliminating the orthorhomic transition that also 
produces the structurally inequivalent dimer bilayers. However, the 
introduction of the smaller Sr ion in the inter-bilayer position of the Ba 
ions also creates a local strain most likely to act along the ${\hat c}$-axis 
within the bilayers. The most important statement concerning any magnetic 
disorder produced by this structural disorder is that it can be at most a 
bond disorder, and not a site disorder, because the Cu$^{2+}$ sites are 
unaffected. Because of the inter-bilayer location of the Ba$^{2+}$ and 
Sr$^{2+}$ ions, this bond disorder is most likely to affect the local 
value of $J^{\prime\prime}$, a parameter so small that our results are not very 
sensitive to the possibility of a distribution of values. The most important 
observation in the context of possible magnetic disorder is the result of our 
INS measurements that any resulting triplon broadening (lifetime) effects are 
too small to be distinguished from extrinsic factors (Sec.~\ref{sins}B). 
From this we may conclude that local magnetic effects arising from this 
disorder are minimal, and that it has no discernible consequences for the 
magnetic Hamiltonian. Thus the disorder can be called ``ideal'' in the sense 
that only its qualitative and quantitative effects on the average structure 
are relevant. Regarding its density, while the nominal substitution was 
$x = 0.1$ and a compatible value of $x = 0.08(2)$ was determined by 
energy-dispersive x-ray spectroscopy (EDX), a somewhat lower value of 
$x = 0.05(1)$ was obtained from Rietveld refinement of synchrotron 
diffraction data for crushed crystals \cite{Puphal2016}. 

Further disorder effects within our experiment are the following. The 
density of magnetic impurities was estimated by fitting the Curie tail in the 
magnetic susceptibility, which arises at the lowest temperatures due to free 
spin-1/2 levels, and was found to be 3\% in the best crystals selected for 
our experiments \cite{Puphal2016,Puphal2019}. A typical nonmagnetic impurity 
in BaCuSi$_{2}$O$_{6}$ is Cu$_{2}$O, which can be created by the decay of CuO 
during crystal growth; it is visible on surfaces and in thinner crystals, 
allowing its presence to be minimized by careful selection of crystals. 
In our INS experiments, the floating-zone-grown single crystals have a 
reorientation in growth direction of up to 15$^{\circ}$ as a consequence of 
the growth process \cite{Puphal2019}. However, it is important to stress that 
none of these possible sources and consequences of disorder are significant 
when compared to the complex and incommensurate crystal structure and magnetic 
Hamiltonian realized in the parent material \cite{Samulon2006,Rueegg2007,
Kraemer2007,Sheptyakov2012,Allenspach2020}. 

\section{Conclusion}
\label{sc}

In summary, we have performed a comprehensive experimental 
and numerical analysis of the quasi-2D dimerized antiferromagnet 
Ba$_{0.9}$Sr$_{0.1}$CuSi$_{2}$O$_{6}$. The non-magnetic site disorder due to 
this weak Sr-substitution of Han Purple affects only the lattice, acting to 
stabilize the tetragonal room-temperature structure of the parent material
down to the lowest temperatures. We have measured the spin excitation spectrum 
at 1.5 K and zero applied magnetic field to verify that indeed there is only 
one triplon branch, and hence that all dimer bilayers are structurally and 
magnetically equivalent in this compound. This triplon has a spin gap of 
3.00 meV and is described by only three Heisenberg superexchange interactions, 
all on very different energy scales. 

To examine the consequences of these interactions for field-induced magnetism 
and universal critical properties at the magnetic transition, we have gathered 
extensive thermodynamic data at high fields and low temperatures using a 
range of experimental techniques, including a detailed NMR study of the 
quantum critical phase boundary. For an optimal analysis of these latter data, 
we have performed Bayesian inference to obtain the critical exponent $\phi
 = 0.67^{+0.05}_{-0.07}$, determined by assuming a quantum critical regime of 
approximate width 0.4 K. QMC simulations of a 3D model with a single type 
of bilayer yielding the triplon band determined from the zero-field INS 
measurements provide excellent agreement with the quantum critical phase 
boundary, return a critical exponent $\phi = 0.66^{+0.01}_{-0.01}$, and indicate 
both a convergence to $\phi = 2/3$ and the actual width of the regime over 
which this quantum critical scaling is obeyed. We conclude that the structural 
alteration caused by the small chemical modification of Han Purple to its 
10\% Sr-substituted variant fully restores the originally envisaged 
properties of 3D quantum critical scaling at the field-induced magnetic 
ordering transition of a structurally and magnetically quasi-2D system.

\begin{acknowledgments}
We thank W. J. Coniglio for technical support. We are grateful to I. Fisher, 
T. Kimura, S. Sebastian, and the late C. Berthier and C. Slichter for their 
foundational contributions to this study. This work is based in part on 
experiments performed at the Swiss Spallation Neutron Source, SINQ, at the 
Paul Scherrer Institute. A portion of the work was performed at the National 
High Magnetic Field Laboratory, which is supported by the National Science 
Foundation Cooperative Agreement No.~DMR-1644779, the State of Florida, and 
the United States Department of Energy. We thank the Swiss National Science 
Foundation and the ERC grant Hyper Quantum Criticality (HyperQC) for financial 
support. Work in Tallinn and travel to large facilities to perform NMR and 
torque measurements were funded by the European Regional Development Fund 
(TK134) and the Estonian Research Council (PRG4, IUT23-7). Work in Toulouse 
was supported by the French National Research Agency (ANR) under projects 
THERMOLOC ANR-16-CE30-0023-02 and GLADYS ANR-19-CE30-0013, and by the use 
of HPC resources from CALMIP (Grant No.~2020-P0677) and GENCI (Grant 
No.~x2020050225). We acknowledge the support of the LNCMI-CNRS, a member 
of the European Magnetic Field Laboratory (EMFL), of the Swiss Data 
Science Centre (SDSC) through project BISTOM C17-12, and of the Deutsche 
Forschungsgemeinschaft (DFG, German Research Foundation) through Grants 
No.~SFB/TR49, UH90/13-1, and UH90/14-1. 
\end{acknowledgments}

\begin{appendix}

\section{Neutron spectrum calculations}
\label{appins}

The single-particle excitations in an AF spin-dimer material with 
a global singlet ground state are coherently propagating triplets, 
commonly referred to as triplons. To determine the interaction parameters of 
Ba$_{0.9}$Sr$_{0.1}$CuSi$_{2}$O$_{6}$, we have modelled the magnetic dynamical 
structure factor, $S(\vec{Q}, \omega)$, by applying a triplon-based RPA 
approach \cite{Leuenberger1984,Allenspach2020}. In this framework, the 
zero-temperature dispersion of the single triplon mode is given by
\begin{equation}
\omega(\vec{Q}) = \sqrt{J^2 + JJ(\vec{Q})}
\end{equation}
for $\vec{Q} \equiv (Q_h, Q_k, Q_l)$ in the crystallographic basis, with
\begin{align}
J(\vec{Q}) = & 2J^{\prime} \big[ \cos(\pi[Q_h + Q_k]) + \cos(\pi[Q_h - Q_k]) 
\big] \\ & - 2J^{\prime\prime} \cos({\textstyle \frac12} \pi Q_l) \big[ \cos( 
\pi Q_h) + \cos(\pi Q_k) \big]. 
\end{align}
The INS intensities, $I(\vec{Q},\omega)$, are given by
\begin{equation}
I(\vec{Q}, \omega) = \mathcal{A} F^{2}_{\rm{mag}}(\vec{Q}) S(\vec{Q},\omega),
\end{equation}
where $\mathcal{A}$ is an overall prefactor. The magnetic form factor, $F_{\rm 
{mag}}(\vec{Q})$, is known to be spatially anisotropic \cite{Allenspach2020}
and hence was modelled using the minimal parametrization
\begin{equation}
F_{\rm{mag}}(\vec{Q}) \! = \! C_1 e^{-(c_{h1}Q_{h}^{2} + c_{l1}Q_{l}^{2})} \! + \! C_2 
e^{-(c_{h2}Q_{h}^{2} + c_{l2}Q_{l}^{2})}.
\end{equation}
Theoretically, the dynamical structure factor at zero temperature due to the 
triplon is given by
\begin{equation}
S(\vec{Q}, \omega) = N \big[ 1 - \cos(\vec{Q} \! \cdot \! \vec{R}) \big] 
\frac{J}{\omega(\vec{Q}\,)} \delta \big(\omega (\vec{Q}) - \omega \big),
\end{equation}
where $N$ is the number of dimers in the crystal and $\vec{R}$ the vector 
connecting the two ions of a single dimer. In experiment, $N$ is absorbed 
in $\mathcal{A}$ and the $\delta$-function is replaced by a Gaussian of 
unit integrated intensity and width $\Gamma(\vec{Q})$, which includes both 
intrinsic dynamical and extrinsic instrumental resolution effects. In the 
modelled spectra shown in Figs.~\ref{fig1} and \ref{fig2}, $\Gamma(\vec{Q})$
was fitted separately at every $\vec{Q}$ point. $\mathcal{A}$ and 
$F^{2}_{\rm{mag}} (\vec{Q})$ were fitted using the intensity data shown in 
Figs.~\ref{fig2}(d) and \ref{fig2}(e) while keeping the interaction parameters 
fixed to their optimal values determined using the dispersion data of 
Figs.~\ref{fig2}(b) and \ref{fig2}(c). 

\section{Extraction of NMR phase boundary}
\label{appnmrpb}

To extract the phase boundary of Ba$_{0.9}$Sr$_{0.1}$CuSi$_{2}$O$_{6}$ from the 
NMR spectra, we modelled the opening and increase of the peak-splitting 
visible in Figs.~\ref{fig4}(a) and \ref{fig4}(b) by exploiting its direct 
proportionality to the order parameter [inset, Fig.~\ref{fig4}(d)]. In the 
critical regime we therefore assumed that the splitting of peak frequencies 
is given by
\begin{equation}
\Delta f(H, T) = c(T) [H - H_c(T)]^{\zeta(T)}
\label{eq:splitting}
\end{equation}
for each measurement temperature, $T$. This allowed us to perform simultaneous 
fits of all spectra at each $T$, and hence to optimize the accuracy of our 
estimate for the corresponding $H_{c}(T)$.

To model the shapes of the NMR peaks [Figs.~\ref{fig4}(a) and \ref{fig4}(b)] we 
used the Student $t$-distribution. A single peak was used below $H_{c}(T)$ and 
two peaks above, where $H_c(T)$ is the key parameter to be fitted. Implementing 
this behavior required a number of additional assumptions. The widths of the 
peaks below $H_c(T)$ [blue in Figs.~\ref{fig4}(a) and \ref{fig4}(b)] and of 
the lower-frequency peaks above $H_c(T)$ (red) were treated as individual 
parameters, while the widths of the higher peaks above $H_c(T)$ (green) 
were parametrized using a low-order polynomial in order to avoid spurious 
contributions from background features. The shape parameters of the 
$t$-distributions were modelled using a function linear in $H - H_c(T)$, 
which allowed us to confirm that the shape remains almost constant for 
different fields. The relative intensity ratio of the two peaks, and the size 
of a constant background parameter, were taken to be the same for all spectra 
measured at the same temperature. To assist the fit convergence, and because 
the important parameter is the position or positions of the NMR peak(s), all 
integrated intensities were normalized to unity before processing. 

\section{Bayesian analysis of phase-boundary data}
\label{appbayes}

To analyze the NMR and QMC data close to the QCP within the framework of 
Eq.~\eqref{eq:phase_boundary}, it is necessary to specify the functions 
entering Eq.~\eqref{eq:BayesTheorem}. Beginning with QMC, the critical 
fields, $H_{c1}$ and $\{H_{c,j}\}$, are input values for the simulation and 
hence are known exactly ($\{\delta H_{c,j}\} = 0$). The corresponding critical 
temperatures, $\{T_{c,j}\}$, are estimates based on extrapolation from repeated 
simulations with increasing grid sizes and we assumed them to obey a normal 
distribution around the true critical fields with standard deviations 
$\{\delta T_{c,j}\}$. The likelihood function was therefore taken to be  
\begin{equation}
p(\mathcal{D}_{\rm QMC} | \bm{\theta}_{\rm QMC}) = \prod_{j=1}^{j_{\mathrm{max}}} 
\mathcal{N} (T_{c,j} | \alpha [H_{c,j} - H_{c1}]^{\phi}, \delta T_{c,j}),
\label{eq:qmcmodel}
\end{equation}
where $\mathcal{N}(x | \mu,\sigma)$ denotes the probability density function 
of the normal distribution with mean $\mu$ and standard deviation $\sigma$, 
evaluated at $x$. In this expression $\bm{\theta}_{\rm QMC} \equiv (\alpha, 
\phi)$ contains only two unknowns and $j_{\rm max}$, the index of the data 
point with temperature $T_{c,j} = T_{\rm max}$, introduces the finite width 
of the quantum critical regime. The statistical relationship between the 
phase-boundary parameters and the observations in this case is therefore 
that of a regression model.

In the phase-boundary data obtained by NMR, both the critical fields and the 
corresponding temperatures are estimated quantities. Each temperature, $T_{c,j}$,
is subject to both measurement uncertainties and possible systematic variations,
and was treated as a noisy measurement of a single value. Each critical field, 
$H_{c,j}$, was obtained from the NMR peak-splitting with an uncertainty that 
arises from background, noise, and the finite number of measurements, as a 
result of which it is essentially uncorrelated with $\delta T_{c,j}$. The 
presence of uncertainty in two measurement dimensions means that the true 
point on the phase boundary at each $j$ is unknown. Its location can be 
parametrized by either a temperature or a field coordinate, and here we chose 
the latter, introducing the parameters $\{H_{c,j}^*\}$ to represent the true, 
unknown critical fields of which $\{H_{c,j}\}$ are uncertain estimates. 
Taking $\{H_{c,j}\}$ and $\{T_{c,j}\}$ to be independent and to have a normal 
distribution around these true phase-boundary points, with respective 
standard deviations $\{\delta H_{c,j}\}$ and $\{\delta T_{c,j}\}$, the 
likelihood function was taken as 
\begin{eqnarray}
p(\mathcal{D}_{\rm NMR} | \bm{\theta}_{\rm NMR}) & = & \prod_{j=1}^{j_{\mathrm{max}}} 
\mathcal{N} (T_{c,j} | \alpha [H_{c,j}^* - H_{c1}]^{\phi}, \delta T_{c,j}) \nonumber 
\\ & & \;\;\;\;\;\;\;\;\;\; \times \mathcal{N} (H_{c,j} | H_{c,j}^*, 
\delta H_{c,j}),
\label{eq:nmrmodel}
\end{eqnarray}
where $\bm{\theta}_{\rm NMR}$ denotes the extended parameter set $(\alpha, H_{c1}, 
\phi, H_{c,j}^*)$. In the Bayesian methodology, once the posterior is obtained 
its dependence on the introduced parameters $\{H_{c,j}^*\}$ can be removed in 
an optimal manner by averaging over all possible values weighted by their 
posterior probability.

Turning to the prior probability distributions in Eq.~\eqref{eq:BayesTheorem}, 
their selection should guide the inference process efficiently without 
introducing undue bias in the final result. $\alpha$ is known from studies 
of BaCuSi$_{2}$O$_{6}$ \cite{Sebastian2006,Allenspach2020} only to be of order 
unity (for measurements in T and K), and hence we assigned it an uninformative 
normal distribution on a logarithmic scale. For $\mu_0 H_{c1}$, we used the gap 
$\Delta = 3.00(1)$~meV obtained from the INS data and the $g$-factor 
$g_{\parallel c} = 2.32(2)$ \cite{Puphal2016} to obtain an initial estimate of 
$22.34 \pm 0.24$ T. Because $\phi$ is positive and is not expected to exceed 
unity, we used as its prior a half-normal distribution, $\mathcal{N}_{\rm{half}}
(x | \sigma)$, with the scale parameter chosen to ensure that it is 
approximately flat in the region where $\phi$ is physically plausible, but 
does permit larger values. These considerations gave the prior probability 
distributions 
\begin{equation}
  \begin{split}
    p(\log_{10} \alpha) &= \mathcal{N} (\log_{10} \alpha | 0, 3), \\
    p(H_{c1}) &= \mathcal{N} (\mu_0 H_{c1} | 22.34~{\rm T}, 0.24~{\rm T}), \\
    p(\phi) &= \mathcal{N}_{\rm{half}} (\phi | 2),
  \end{split}
  \label{eq:prior}
\end{equation}
for the three model parameters. For the true critical fields, $\{H_{c,j}^*\}$, 
required to analyze the NMR phase boundary, these are unknown before 
measurement and thus we assigned a flat prior probability distribution,
\begin{equation}
p(H_{c,j}^*) = \mathcal{U}(\mu_0 H_{c,j}^* | 0\,\mathrm{T}, 50\,\mathrm{T}),
\end{equation}
where $\mathcal{U}(x | a,b)$ denotes a uniform distribution with limits $a$ 
and $b$. We made the minimal assumption that the range of possible values does 
not exceed 0-50 T, which includes the entirety of the ordered phase. Because 
all of these probabilities are independent, the joint prior probability 
distributions, $p(\bm{\theta}_{\rm QMC})$ and $p(\bm{\theta}_{\rm NMR})$, entering 
Eq.~(\ref{eq:BayesTheorem}) are given by their product. 

To investigate the unknown width of the quantum critical regime, we repeated 
the analysis of our NMR data multiple times with $339~{\rm mK} \le T_{\rm max} 
\le 921~{\rm mK}$ ($5 \le j_{\rm max} \le 10$ data points) and of our QMC data 
with $234~{\rm mK} \le T_{\rm max} \le 939~{\rm mK}$ ($8 \le j_{\rm max} \le 30$ 
data points). In each case we recovered the posterior by drawing random 
samples using \textit{emcee} \cite{ForemanMackey2013}, a Markov-chain Monte 
Carlo implementation of an affine invariant ensemble sampler \cite{Goodman2010}.
An ensemble of 200 chains was sampled for 40\,000 steps to ensure equilibration 
of the Markov process before a further 10\,000 steps were recorded, yielding a 
total of 2\,000\,000 parameter samples for each $j_{\rm max}$. These samples form 
point clouds in the parameter space of $\bm{\theta}$, whose density is 
proportional to the posterior probability, and one of these (obtained using 
the NMR data with $j_{\rm max} = 6$) is visualized using different projections in 
Figs.~\ref{fig5}(a)-\ref{fig5}(d). As a heuristic confirmation of convergence 
we note that the maximal autocorrelation time \cite{Goodman2010} was 117 steps. 
The posterior standard deviations are at least 20 times smaller than the prior 
ones in all cases, confirming that, after convergence, the results depend only 
on the data and not on our choice of weakly informative prior.

\end{appendix}


\end{document}